\documentclass[aps, pre, a4paper, twocolumn, longbibliography, nofootinbib]{revtex4-2}

\usepackage[utf8]{inputenc}
\usepackage{epsfig}
\usepackage[T1]{fontenc}
\usepackage[english]{babel}
\usepackage[table]{xcolor}
\usepackage{t1enc}
\usepackage{graphicx}
\usepackage{amssymb}
\usepackage{amsmath}
\usepackage{relsize}
\usepackage[percent]{overpic}
\usepackage{bm}
\usepackage{hyperref}
\usepackage{float}

\usepackage{siunitx}
\usepackage{accents}
\usepackage{mathtools}

\usepackage{grffile}

\usepackage[normalem]{ulem}

\usepackage{pgf}
\usepackage{tikz}
\usetikzlibrary{calc}

\usepackage{longtable}

\newcommand{\dd}{\mathrm{d}}

\usepackage{mathtools}
\def\multiset#1#2{\ensuremath{\left(\kern-.3em\left(\genfrac{}{}{0pt}{}{#1}{#2}\right)\kern-.3em\right)}}

\usepackage{amsmath}

\newcommand{\A}{\bm{A}}

\newcommand{\bb}{\bm{b}}

\newcolumntype{P}[1]{>{\centering\arraybackslash}p{#1}}

\usepackage{verbatim}
\usepackage{overpic}
\usepackage{booktabs}
\usepackage{placeins}

\usepackage{siunitx}
\usepackage{multirow}

\usepackage{pgfplots}

\providecommand\mathplus{+}

\begin{document}

\title{Ordered community detection in directed networks}

\author{Tiago P. Peixoto}
\email{peixotot@ceu.edu}
\affiliation{Department of Network and Data Science, Central European University, 1100 Vienna, Austria}

\begin{abstract}
  We develop a method to infer community structure in directed networks
  where the groups are ordered in a latent one-dimensional hierarchy
  that determines the preferred edge direction. Our nonparametric
  Bayesian approach is based on a modification of the stochastic block
  model (SBM), which can take advantage of rank alignment and coherence
  to produce parsimonious descriptions of networks that combine ordered
  hierarchies with arbitrary mixing patterns between groups. Since our
  model also includes directed degree correction, we can use it to
  distinguish non-local hierarchical structure from local in- and
  out-degree imbalance---thus removing a source of conflation present in
  most ranking methods. We also demonstrate how we can reliably compare
  with the results obtained with the unordered SBM variant to determine
  whether a hierarchical ordering is statistically warranted in the
  first place. We illustrate the application of our method on a wide
  variety of empirical networks across several domains.
\end{abstract}

\maketitle
%\tableofcontents

\section{Introduction}

Interacting entities in a variety of networked systems form pairwise
relationships that are not necessarily symmetric, i.e. an interaction
from $i$ to $j$ is distinct from one from $j$ to $i$. Typical examples
are predator-prey relationships in food
webs~\cite{williams_simple_2000}, antagonist animal
behavior~\cite{strauss_centennial_2022}, reported friendships in social
networks~\cite{rapoport_study_1961}, and the synaptic connection between
neurons. In many such systems, it is often posited that the preferred
direction of interaction can be ascribed to an unobserved ordering of
the elements involved---placing them on a strict one-dimensional latent
hierarchy that most relationships tend to respect. Prominent examples of
such ordered systems are species taxa in food
webs~\cite{johnson_trophic_2014}, and dominance hierarchies in animal
societies~\cite{strauss_centennial_2022}.

However, even when present, directed hierarchies are rarely the only
dimension that determines how interactions take place. For example,
regardless of direction, connections can occur preferentially between
specific types of entities, resulting in compartmentalization and
heterogeneous mixing patterns that are independent of any underlying
ordering. Furthermore, it is also possible for the directed structure of
a network not to be associated with any latent hierarchy at all, and to
be due instead to entirely different mechanisms. Although in such
situations it may still be possible to order the nodes in such a way
that the majority of interactions end up respecting a seeming hierarchy,
this does not necessarily mean that this is in fact a plausible
explanation for how the directions were chosen.

In this work we present a method to infer the \emph{ordered} modular
structure of networks in a manner that simultaneously captures arbitrary
mixing patterns and directed hierarchies. Our method is based on a
modification of the directed version of the stochastic block model
(SBM)~\cite{holland_stochastic_1983,peixoto_bayesian_2019}---a
generative model that can capture arbitrary preferences between groups
of nodes. In our modification, the groups themselves are ordered, such
that the preferred direction of interaction tends to obey their ranking,
while still allowing for the groups to be connected in arbitrary ways,
independent of direction. One important ingredient of our model is
directed degree-correction~\cite{karrer_stochastic_2011}, which allows
nodes that belong to the same group/rank to possess an arbitrarily
varied number of incoming and out-going connections. This means that our
method is capable of distinguishing between merely local
asymmetries---that stem solely from a node's tendency to have a
particular balance of in and out-connections---and actual hierarchies
that affect the structure of the network at a larger scale.

In our methodology we exploit the formal equivalence between statistical
inference and data
compression~\cite{rissanen_modeling_1978,grunwald_minimum_2007,peixoto_bayesian_2019}. In
this setting, we seek to obtain the model inference with the optimal
balance between quality of fit and model complexity, such that the
amount of information required to describe the network is
minimized. This amounts to a nonparametric Bayesian method that can
not only determine in a principled manner the most appropriate number of
ordered groups, but it also allows us to decide whether a hierarchical
structure is warranted at all in the first place, or if we have more
evidence instead for a model alternative without any particular ordering
between the nodes, but which happens to be more compressive.

Our approach can be compared to previous work in the literature in some
important ways. There are several methods that extract relative rankings
between the nodes of a network, based on spectral node
centrality~\cite{page_pagerank_1999,bonacich_power_1987,fogel_serialrank_2014,cucuringu_sync-rank_2016},
minimum violation
ranking~\cite{ali_minimum_1986,slater_inconsistencies_1961,gupte_finding_2011,jiang_statistical_2011,cantwell_belief_2021},
random utility
models~\cite{bradley_rank_1952,luce_possible_1959,david_ranking_1987},
and latent space
models~\cite{williams_probabilistic_2011,ball_friendship_2013,de_bacco_physical_2018,kawamoto_sequential_2021}. The
most central difference between these methods and the one presented in
this work is that none of them attempt to simultaneously detect
community structure, or include degree-correction. Furthermore, with the
exception of the latent space models, these approaches do not attempt to
model the placement of the edges, only their latent
ordering. Additionally, since they do not attempt to make a statement
about data generative processes, they cannot articulate the notion of
statistical significance or parsimony~\cite{peixoto_descriptive_2022}.

The works that are perhaps closest to ours are the approaches from
Letizia et al~\cite{letizia_resolution_2018} and Iacovissi et
al~\cite{iacovissi_interplay_2021}. Letizia et
al~\cite{letizia_resolution_2018} considered a ranked SBM with uniform
connection probabilities between groups depending only on whether the
edge direction violates or not the hierarchy. Besides being unable to
uncover heterogeneous mixing patterns and lacking degree correction, the
approach of Ref.~\cite{letizia_resolution_2018} is not based on a model
likelihood, and hence cannot be used to evaluate statistical
evidence. The method of Iacovissi et al~\cite{iacovissi_interplay_2021}
is based on a different idea, and combines the SBM with
Springrank~\cite{de_bacco_physical_2018}, such that a node can
\emph{either} have a group membership \emph{or} a ranking, but not both
simultaneously. Their model not only lacks degree correction, but its
inference is performed in a parametric fashion: the number of groups in
the SBM needs to be set a priori, and cannot be extracted from the data
itself. Furthermore, the inference procedure developed in
Ref.~\cite{iacovissi_interplay_2021} is based on a variational
approximation, whereas our approach is based on MCMC using an exact
likelihood.

This work is organized as follows. In Sec.~\ref{sec:model} we describe
the model and its inference, and in Sec.~\ref{sec:preference} we
demonstrate how it can be used to simultaneously uncover connection
preference and ranking. In Sec.~\ref{sec:deg-corr} we investigate the
role of degree-correction in distinguishing local from global ordering,
and in Sec.~\ref{sec:model-selection} we consider the problem of model
selection between alternatives without latent ordering. We finalize in
Sec.~\ref{sec:conclusion} with a conclusion.

\section{Network compression via modular structure, rank coherence and alignment}\label{sec:model}

We begin by reviewing how the arbitrary mixing pattern between groups of
nodes of a directed network can be modelled by the microcanonical
degree-corrected stochastic block model
(DC-SBM)~\cite{peixoto_nonparametric_2017}. In this model, the $N$
nodes are divided into $B$ groups, according to a labelled partition
$\bb=\{b_i\}$, where $b_i\in [0,B-1]$ is the group membership of node
$i$. As an additional set of parameters, we have the group affinity
matrix $\bm e =\{e_{rs}\}$, where $e_{rs}$ is the number of directed
edges that are allowed to exist from group $s$ to $r$, as well of
the out-/in-degree sequence $\bm k =
\{(k^{\text{out}}_i,k^{\text{in}}_i)\}$, where $k^{\text{out}}_i$ and
$k^{\text{in}}_i$ are the out- and in-degrees of node $i$,
respectively. With these constraints in place, a directed multigraph
$\bm A = \{A_{ij}\}$, where $A_{ij}$ is the number of edges from $j$ to
$i$, is generated by placing $k^{\text{out}}_i$ and $k^{\text{in}}_i$
``half-edges'' on each node $i$, and then pairing them uniformly at
random while respecting the counts $e_{rs}$ between all groups $r$ and
$s$. A resulting multigraph $\bm A$ is sampled in this manner with
probability~\cite{peixoto_nonparametric_2017}
\begin{equation}\label{eq:dc-sbm}
  P(\A|\bm k, \bm e, \bm b) =
  \frac{\prod_{rs}e_{rs}!\prod_ik_i^{\text{out}}!k_i^{\text{in}}!}
  {\prod_{ij}A_{ij}!\prod_re_r^{\text{out}}!e_r^{\text{in}}!},
\end{equation}
with $e_r^{\text{out}}=\sum_se_{sr}$ and $e_r^{\text{in}}=\sum_se_{rs}$,
as long as the imposed constraints are respected, otherwise the
probability is zero.\footnote{It is possible derive our approach in an
entirely equivalent manner by replacing Eq.~\ref{eq:dc-sbm} with
independent Poisson distributions for each entry $A_{ij}$, and
marginalizing over their parameters~\cite{peixoto_nonparametric_2017},
but the microcanonical formulation is more convenient for our purposes.}

The task of identifying the most plausible division of a directed
network $\A$ into groups consists in inverting the above procedure, and
obtaining the posterior distribution
\begin{equation}\label{eq:posterior}
  P(\bb | \A) = \frac{P(\A|\bb) P(\bb)}{P(\A)},
\end{equation}
where $P(\bb)$ is the prior for the node partition, and $P(\A|\bb)$ is
the marginal likelihood,
\begin{align}
  P(\A|\bb)
  &= \sum_{\bm k, \bm e}P(\A|\bm k, \bm e, \bm b)P(\bm k, \bm e| \bm b)\\
  &= P(\A|\hat{\bm k}, \hat{\bm e}, \bm b)P(\hat{\bm k}, \hat{\bm e}| \bm b),
\end{align}
where $\hat{\bm k}$ and $\hat{\bm e}$ are the only parameter values
compatible with the network $\A$ and partition $\bb$. The prior $P(\bm
k,\bm e,\bb)$ is derived in Ref.~\cite{peixoto_nonparametric_2017} and
described in Appendix~\ref{app:priors} for completeness. Finding the
partition $\bb$ that maximizes Eq.~\ref{eq:posterior} is equivalent to
minimizing the \emph{description length} of the
model~\cite{grunwald_minimum_2007}, given by
\begin{align}\label{eq:dl}
  \Sigma(\A,\bb) = -\log_2P(\A|\hat{\bm k}, \hat{\bm e}, \bm
  b)-\log_2P(\hat{\bm k}, \hat{\bm e}, \bm b).
\end{align}
The first term in the right hand side of above equation determines the
minimum length of a binary message that is required to transmit the
matrix $\A$, in such a manner that it can be decoded from the message
without errors, provided the parameter values $\hat{\bm k}$, $\hat{\bm
e}$ and $\bb$ are already known by the receiver. Likewise, the second
term determines the amount of information needed to transmit the model
parameters themselves. Therefore, the resulting value $\Sigma(\A,\bb)$
corresponds to the total length of the shortest message that is required
to transmit the network $\A$ to a receiver that has no prior information
on its structure, which must involve sending the parameter values as
well.

Minimizing the description length $\Sigma(\A,\bb)$ has the desirable
effect of preventing \emph{overfitting}, which happens for example when
we choose a number of groups $B$ that is too large, and the inferred
modular structure captures spurious random
fluctuations~\cite{guimera_modularity_2004}. This is because if a
portion of the network (or its entirety) has been generated by a
maximally random placement of the edges, it becomes asymptotically
impossible to compress it with any algorithm---maximally random data are
inherently \emph{incompressible}~\cite{cover_elements_1991}. Therefore,
if splitting a set of nodes into two groups significantly reduces the
description length, this means that the placement of the edges involved
is very unlikely to have been maximally random, and hence the division
is capturing statistically significant structure.

More operationally, the second term in right hand side of Eq.~\ref{eq:dl}
serves as a \emph{penalty} to the first term, since it tends to increase
 together with the model complexity, while the first term tends to
decrease as the larger number of constraints match the data more
closely. The optimal inference is therefore a balance between these two
aspects---model complexity and quality of fit---and the overall method
serves as formal implementation of Occam's razor (or the principle of
parsimony), which states that simpler models are preferable to more
complex ones, provided they have the same explanatory power.

With the posterior of Eq.~\ref{eq:posterior} in place, we can proceed in
two ways, depending on our objective. We can find the single partition
$\bb$ that maximizes that probability, which also minimizes the
description length. Alternatively, we can sample partitions from this
distribution, and in this way explore the entire landscape of
hypotheses, weighted according to their plausibility. The latter can
also be seen as a minimum description length (MDL) scheme, with a ``one
part'' description length given by the full marginal distribution,
i.e. $\Sigma(\A)=-\log_2 P(\A)$, with $P(\A)=\sum_{\bb}P(\A|\bb)P(\bb)$,
and noting that $\Sigma(\A)\le \Sigma(\A,\bb)$, i.e. the full marginal
description length is strictly shorter or equal to the one conditioned
on a single partition. Both tasks can be accomplished efficiently using
Markov chain Monte-Carlo (MCMC), as described in
Refs.~\cite{peixoto_efficient_2014,peixoto_merge-split_2020}.

\subsection{Identifying group orderings}

Although the above model is capable of uncovering directed preferences
between groups of nodes, including those where an underlying ordering is
present, the ordering itself is not revealed by the model
parameters. This is because the posterior distribution of
Eq.~\ref{eq:posterior}---and therefore also the description length of
Eq.~\ref{eq:dl}---is invariant to permutations of the group labels. More
specifically, if we consider two partitions $\bb$ and $\bm c$, such that
\begin{equation}
  b_i = \mu(c_i),
\end{equation}
where $\mu(r)$ is a bijection of the group labels, then we have
\begin{equation}
  P(\bb|\A) = P(\bm c|\A),\quad \Sigma(\A,\bb) = \Sigma(\A,\bm c).
\end{equation}
Therefore, the ordering of the groups is entirely immaterial and cannot
be used to attain compression under this model, and reveal any aspect of
the network structure.

Here we modify precisely this property of the model via a relatively
simple, but consequential change. In fact, we keep the model of
Eq.~\ref{eq:dc-sbm} exactly as it is, together with the priors for $\bm
k$ and $\bb$, and we change only the prior for the group affinities,
$\bm e$. First, we introduce the auxiliary parameter $m_{rs}$, which
counts the total number of edges between groups $r$ and $s$ (or twice
that number if $r=s$), regardless of edge direction, i.e.
\begin{align}
  m_{rs} &= \sum_{ij}(A_{ij}+A_{ji})\delta_{b_i,r}\delta_{b_j,s}.
\end{align}
Conditioned on this number, we sample the upstream ($e_{rs}$, with
$r>s$) and downstream ($e_{rs}$, with $r<s$) affinities according to
\begin{equation}
  P(e_{rs},e_{sr} | m_{rs}, p) =
  \begin{cases}
    \delta_{e_{sr},m_{rs}-e_{rs}}P(e_{rs} | m_{rs}, p) & \text{ if } r<s,\\
    \delta_{e_{rs},m_{rs}-e_{sr}}P(e_{sr} | m_{rs}, p) & \text{ if } r>s,
  \end{cases}
\end{equation}
ensuring that $e_{rs} + e_{sr}=m_{rs}$, and with the downstream affinity
sampled according to a binomial distribution with parameter $p$,
\begin{equation}
  P(e_{rs} | m_{rs}, p) =
  {m_{rs}\choose e_{rs}}p^{e_{rs}}(1-p)^{m_{rs}-e_{rs}}.
\end{equation}
We call edges that connect nodes of the same group as ``lateral,'' since
they go neither upstream nor downstream. The lateral
affinities are given directly by $\bm m$,
\begin{equation}
  P(e_{rr} | m_{rr}) = \delta_{e_{rr},m_{rr}/2}.
\end{equation}
Introducing the total number of upstream, downstream, and lateral edges,
\begin{equation}
  E^+ = \sum_{r<s}e_{sr},\quad
  E^- = \sum_{r<s}e_{rs},\quad
  E^0 = \sum_{r}e_{rr},
\end{equation}
respectively, allows us to write the total conditional probability,
\begin{equation}
  P(\bm e | \bm m, p) = \left[\prod_{r<s}{m_{rs}\choose e_{rs}}\right] p^{E^{-}}(1-p)^{E^{+}}.
\end{equation}
The parameter $p$ is considered to be unknown \emph{a priori}, so we
compute the marginal probability,
\begin{align}
  P(\bm e | \bm m) &= \int_0^1 P(\bm e | \bm m, p)P(p)\,\dd p\\
  &= \left[\prod_{r<s}{m_{rs}\choose e_{rs}}\right]
  {E^{+}+E^{-}\choose E^{+}}^{-1}\\
  &\quad\times\frac{1}{E^{+}+E^{-}+1},
\end{align}
where we have used a uniform prior density $P(p)=1$. For the symmetric
matrix $\bm m$, we use a uniform distribution conditioned on the total
number of edges $E=E^++E^-+E^0$, given by
\begin{align}\label{eq:prior_m}
  P(\bm m|E, B) = \multiset{\multiset{B}{2}}{E}^{-1},
\end{align}
where $\multiset{n}{m}={n+m-1\choose m}$ is the number of
$m$-combinations from a set of size $n$, allowing for repetitions.
Putting all together, we have
\begin{align}
  P(\bm e | E, B) &= \left[\prod_{r<s}{e_{rs} + e_{sr}\choose e_{rs}}\right]
  {E^{+}+E^{-}\choose E^{+}}^{-1}\nonumber\\
  &\quad\times\frac{1}{E^{+}+E^{-}+1}\times \multiset{\multiset{B}{2}}{E}^{-1}.
\end{align}
Since this probability will depend on the overall number of downstream,
upstream, and lateral edges, the resulting description length will no
longer be invariant to arbitrary label permutations. However, it is
still invariant to full rank \emph{reversals}, i.e. the specific group
label bijection $\mu(r) = B-1-r$, which would cause an overall reversal
of the upstream and downstream directions. Therefore, the overall
top-down or down-top orientation of the ordering is not identifiable with
this model---but this is hardly relevant in most contexts, since we are
interested only in relative rankings. Without loss of generality, for
presentation purposes we will adopt the convention that most edges
always flow upstream, i.e. $E^+\ge E^-$, since a result obtained with
the opposite flow can always be reversed without changing the
description length.

\begin{figure}
  \begin{tikzpicture}
    \begin{axis}[ymin=0,ymax=1,xmax=1,xmin=0,
        xlabel=Rank alignment,
        ylabel=Rank coherence,
        axis x line=top,
        axis y line=left,
        %axis equal=true,
        ticks=none,
        y dir=reverse,
        width=\columnwidth,
        height=\columnwidth,
        x label style={at={(axis description cs:0.5,1)},anchor=north},
        y label style={at={(axis description cs:0.12,.5)}}]

      \node at (49.5,50.5) {\includegraphics[width=.83\columnwidth]{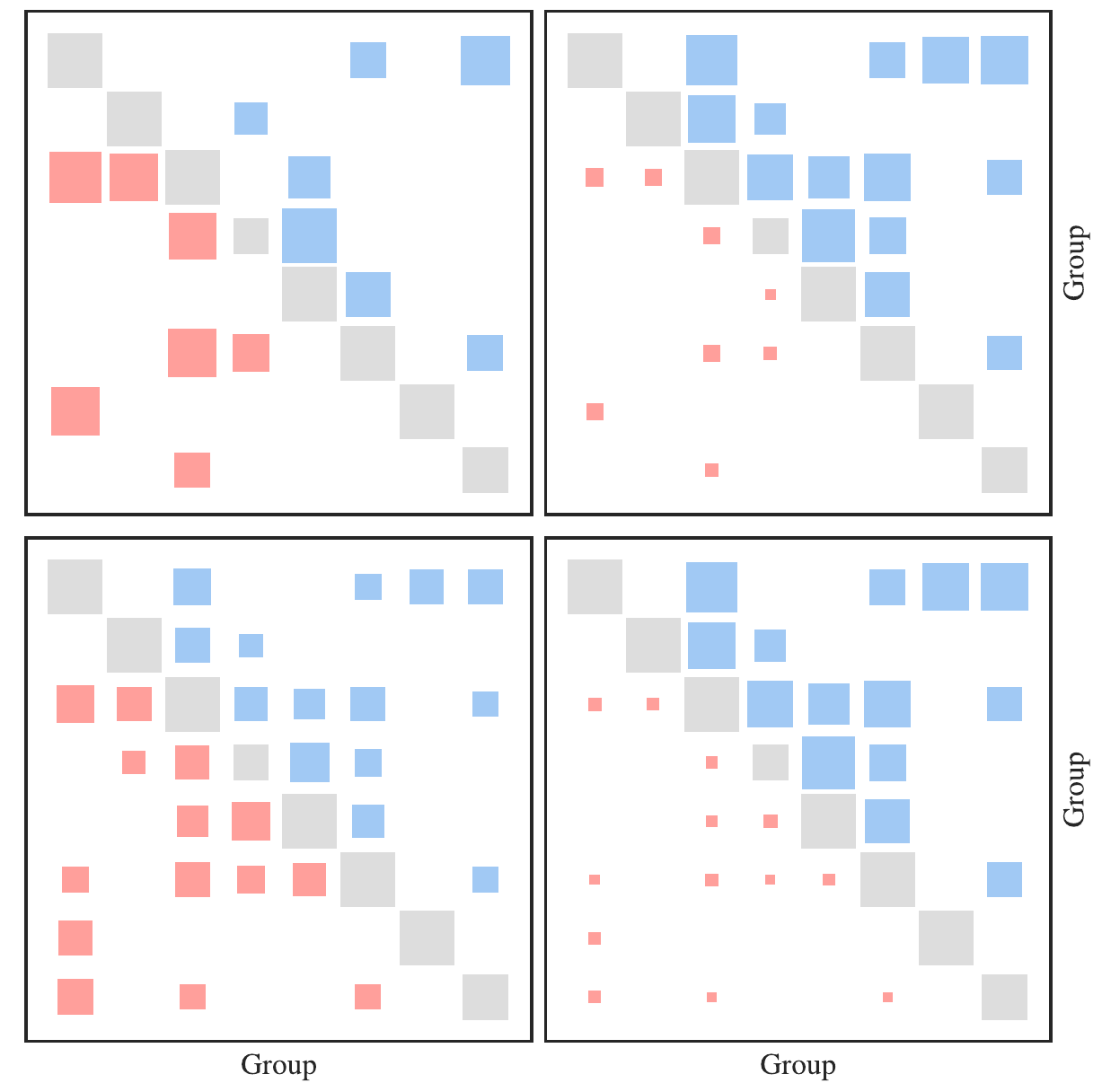}};
  \end{axis}
  \end{tikzpicture}
  \caption{Illustration of the properties of rank alignment and
  coherence. Each panel shows an affinity matrix $e_{rs}$, with upstream
  entries ($r>s$) shown in blue, downstream entries ($r<s$) shown in
  red, and lateral entries ($r=s$) shown in grey, and area of the square
  corresponding to the entry magnitude. High rank alignment means an
  overall abundance of upstream edges, whereas high rank coherence means
  an overall uniformity of pairwise alignments,
  $\Delta_{rs}=e_{rs}-e_{sr}$.\label{fig:coherence}}
\end{figure}

This model formulation can exploit latent orderings as an opportunity
for compression, via the contribution to the description length given by
$\Sigma(\bm e)=-\log_2P(\bm e|E,B)$. There are two different properties
that can make this possible, which we describe in turn: rank alignment
and rank coherence, as illustrated in Fig.~\ref{fig:coherence}.

The local rank alignment between two groups $r$ and $s$, with $r>s$, is
simply by the difference between upstream and downstream affinities,
\begin{equation}
  \Delta_{rs} = e_{rs} - e_{sr}.
\end{equation}
The overall rank alignment is then simply,
\begin{equation}
  \Delta = \sum_{r>s}\Delta_{rs} = E^+ - E^-.
\end{equation}
The larger the magnitude of the overall alignment $\Delta$, the shortest
will be the description length. We can see this by writing the
contribution to the description length as
\begin{multline}\label{eq:dla}
  \Sigma(\bm e) = -\sum_{r>s}\log_2{m_{rs}\choose \frac{m_{rs}+\Delta_{rs}}{2}} + \log_2 {E - E^0 \choose \frac{E - E^0 + \Delta}{2}}\\
  + \log_2(E - E^0 +1) +\log_2 \multiset{\multiset{B}{2}}{E},
\end{multline}
where we use the shorthand $m_{rs}=e_{rs} + e_{sr}$.  The maximal rank
alignment, $\Delta=E-E^0$, achieved with $\Delta_{rs}=m_{rs}$, will
result in the smallest possible description length contribution,
\begin{equation}
  \Sigma(\bm e) = \log_2 \multiset{\multiset{B}{2}}{E} + \log_2(E-E^0+1),
\end{equation}
for fixed values of $B$, $E$, and $E^0$.

Rank coherence, on the other hand, is
the uniformity of the values of $\Delta_{rs}$ across all pairs
$(r,s)$. Maximal rank coherence is when all pairwise rank alignments
coincide with the overall alignment, i.e.
\begin{equation}
  \Delta_{rs} = \frac{\Delta}{E-E^0} \times m_{rs}, \quad \forall\; r > s.
\end{equation}
This results in the first term of right hand side of Eq.~\ref{eq:dla}
given by
\begin{equation}
  -\sum_{r<s}\log_2{m_{rs}\choose \frac{\Delta+E-E^0}{2(E-E^0)}m_{rs}}.
\end{equation}
This is the smallest value this term can take, for fixed $\Delta$ and
$\bm m$ values. Conversely, minimal rank coherence is when the values of
$\Delta_{rs}$ are distributed only between their maximum and minimum
values for different $(r,s)$, i.e. $\Delta_{rs}\in\{m_{rs},
-m_{rs}\}$. In this case, the first term will vanish completely from the
right hand side of Eq.~\ref{eq:dla}, yielding in a strictly larger
description length contribution, if the overall rank alignment $\Delta$
stays the same. Therefore, rank coherence will always provide improved
compression for fixed $\Delta$ and $\bm m$ values.

From the above, we can conclude that when rank alignment is maximal,
rank coherence must also be maximal, and therefore it amounts for the
largest compression possible under this scheme. For intermediary
alignment, a range of rank coherence is allowed, with a larger coherence
providing better compression.

\begin{figure}
  \includegraphics[width=\columnwidth]{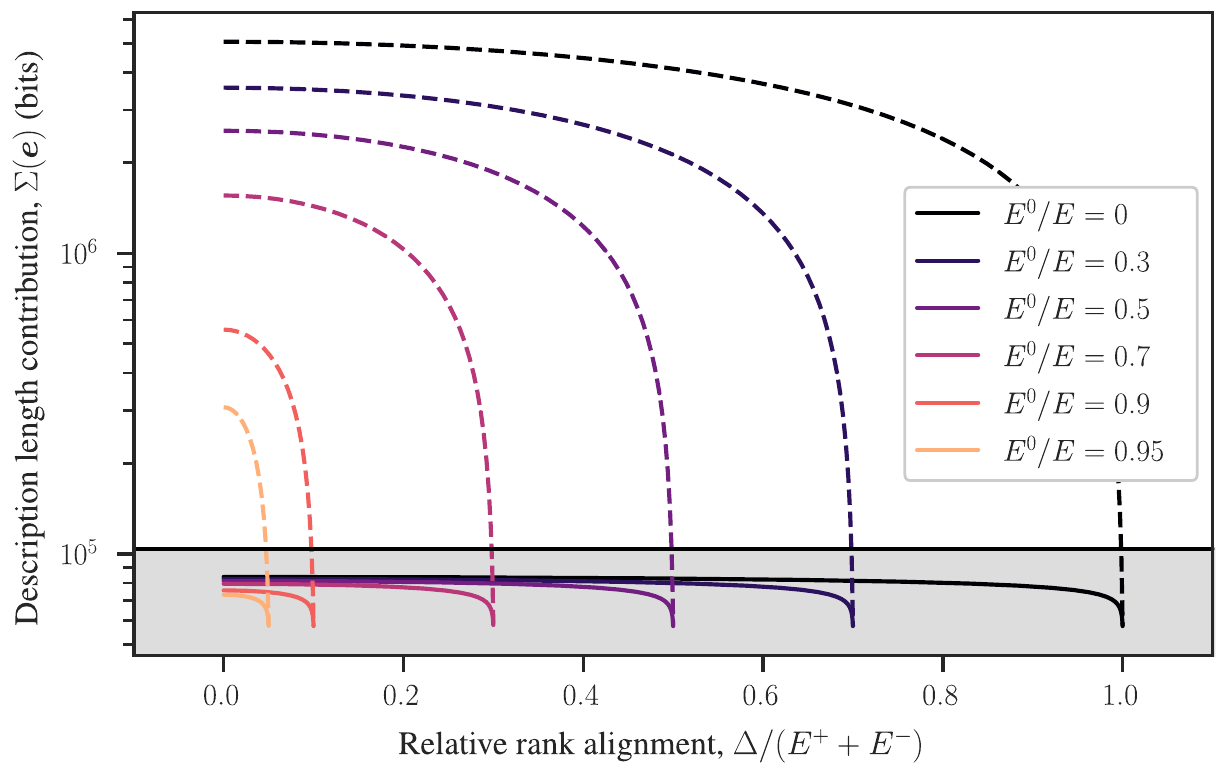} \caption{Description
  length contribution $\Sigma(\bm e)$, as a function of the rank
  alignment $\Delta/(E^++E^-)$, considering both maximal (solid lines)
  and minimal (dashed lines) rank coherence, for different fractions of
  lateral edges (as indicated by the legend), and a value of $E=5\times
  10^6$ and $B=100$. The solid horizontal line marks the value
  $\log_2\multiset{B^2}{E}$ given by Eq.~\ref{eq:dl_r}, and the shaded
  region below it corresponds to a relative compression of the ordered
  parametrization.\label{fig:prior_comp}}
\end{figure}

To understand better the compression that is achievable with group
ordering, it is useful to compare the above prior with the original
uniform choice of the DC-SBM, where the asymmetric matrix $\bm e$ is
sampled directly from a uniform distribution,
\begin{align}
  P'(\bm e|E, B) = \multiset{B^2}{E}^{-1}.\label{eq:dl_r}
\end{align}
With this original choice we recover group label invariance, and hence
cannot profit from any compressibility associated with latent group
orderings. In Fig.~\ref{fig:prior_comp} we compare Eq.~\ref{eq:dla} with
Eq.~\ref{eq:dl_r}, as a function of rank alignment, both for maximum and
minimum rank coherence. As we can see, maximal rank coherence can
achieve better compression than the uniform distribution independent of
the rank alignment magnitude. This means that even when the rank
alignment is zero, with $e_{rs}=e_{sr}$ for every group pair $(r,s)$, we
nevertheless have a more parsimonious explanation of the data using this
model. (This is understandable, since for the matrix $\bm e$ is
symmetric in this situation, which is a kind of structure that cannot be
exploited by the model Eq.~\ref{eq:dl_r} to achieve compression.)

However, if the rank coherence is sufficiently decreased, then the
ordered model no longer offers improved compression over the uniform
distribution of Eq.~\ref{eq:dl_r}. In this situation, the rank
violations become so heterogeneous, that it becomes no longer
parsimonious to describe the group affinities via a group ordering, even
if a majority of edges go in the same direction---we are better off
simply abandoning the ordering altogether, and describing the matrix
$\bm e$ according to arbitrary group labels.

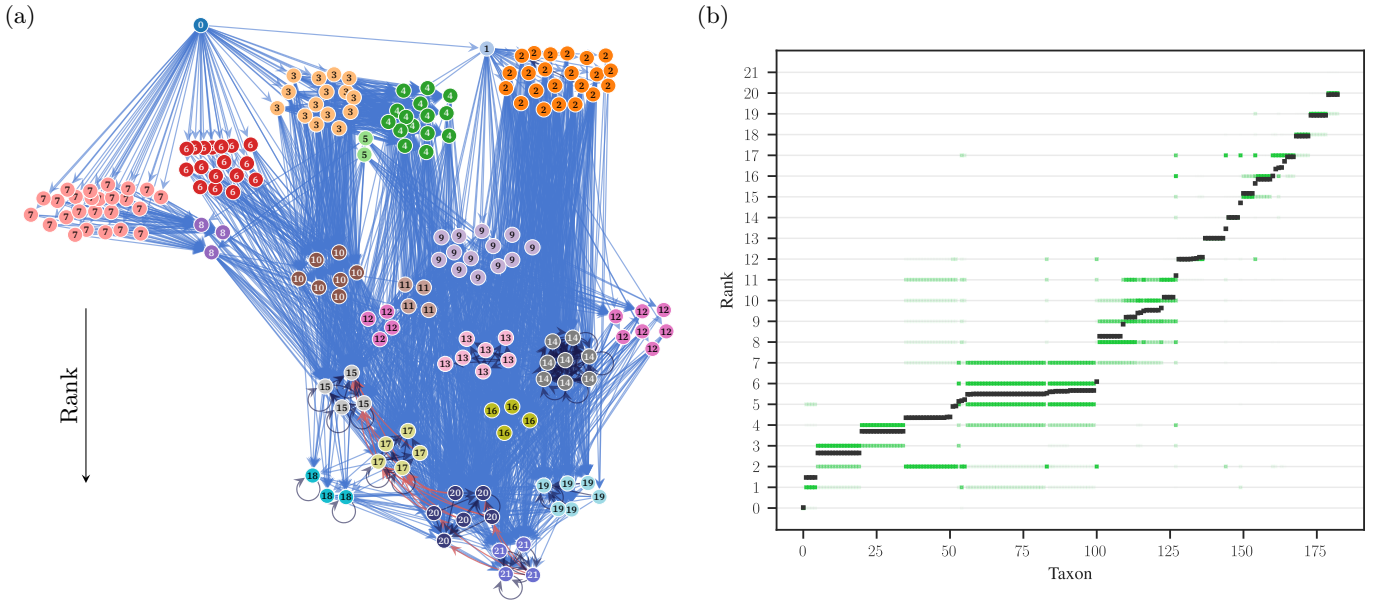
\begin{figure*}[t!]
  \begin{tabular}{cc}
    \raisebox{-\height}{\begin{tikzpicture}
      \node [above right, inner sep=0] (image) at (0,0) {\includegraphics[width=\columnwidth, trim=1cm 1cm 1cm 1cm]{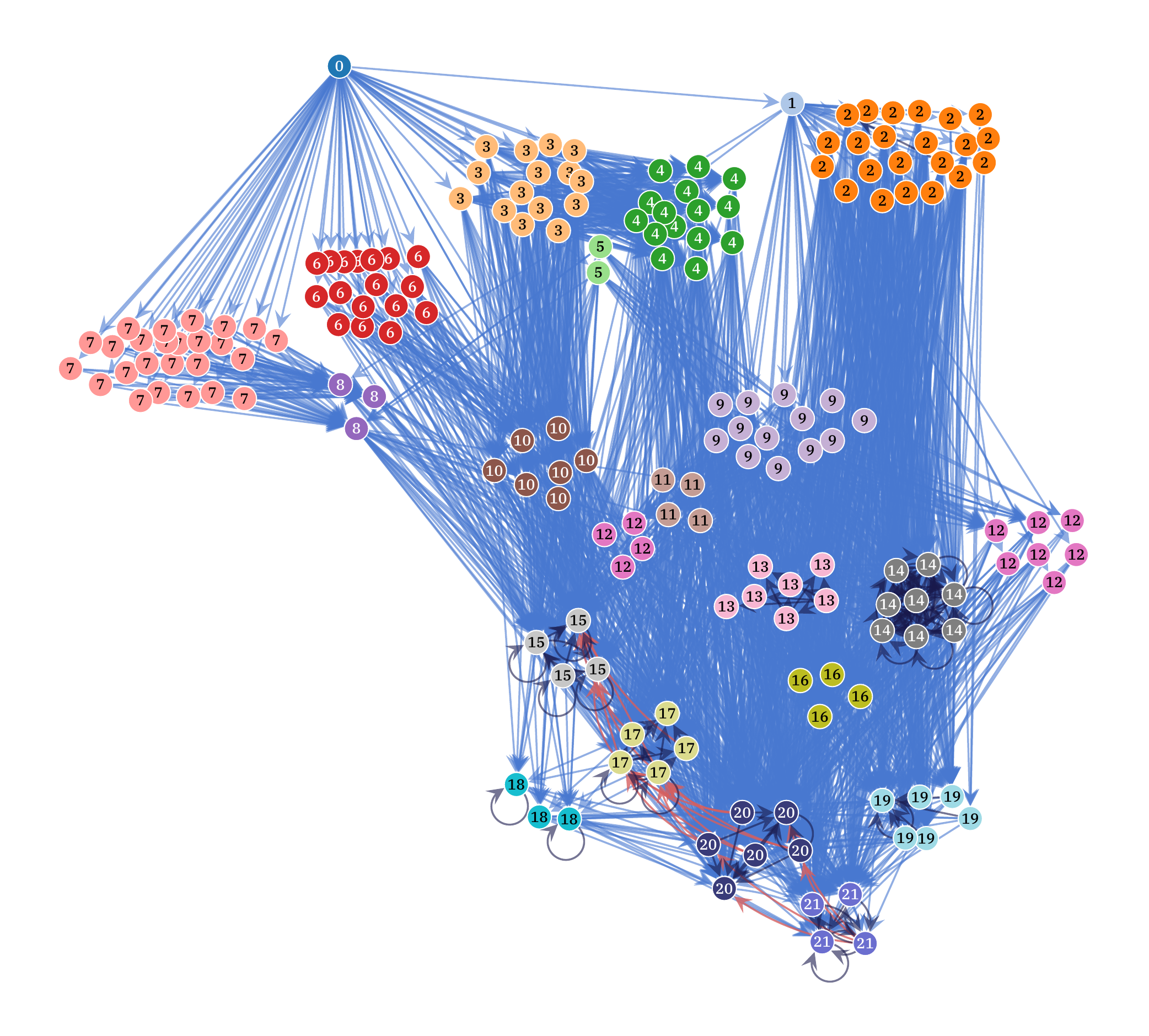}};
      \begin{scope}[
          x={($0.1*(image.south east)$)},
          y={($0.1*(image.north west)$)}]
        \draw [-stealth](1,5) --  node [above,rotate=90] {Rank} (1,2);
        \node at (0, 10) {(a)};
      \end{scope}
    \end{tikzpicture}}&
    \raisebox{-\height}{\begin{tikzpicture}
      \node [above right, inner sep=0] (image) at (0,0) {\includegraphics[width=\columnwidth, trim=0 0 0 -.5cm]{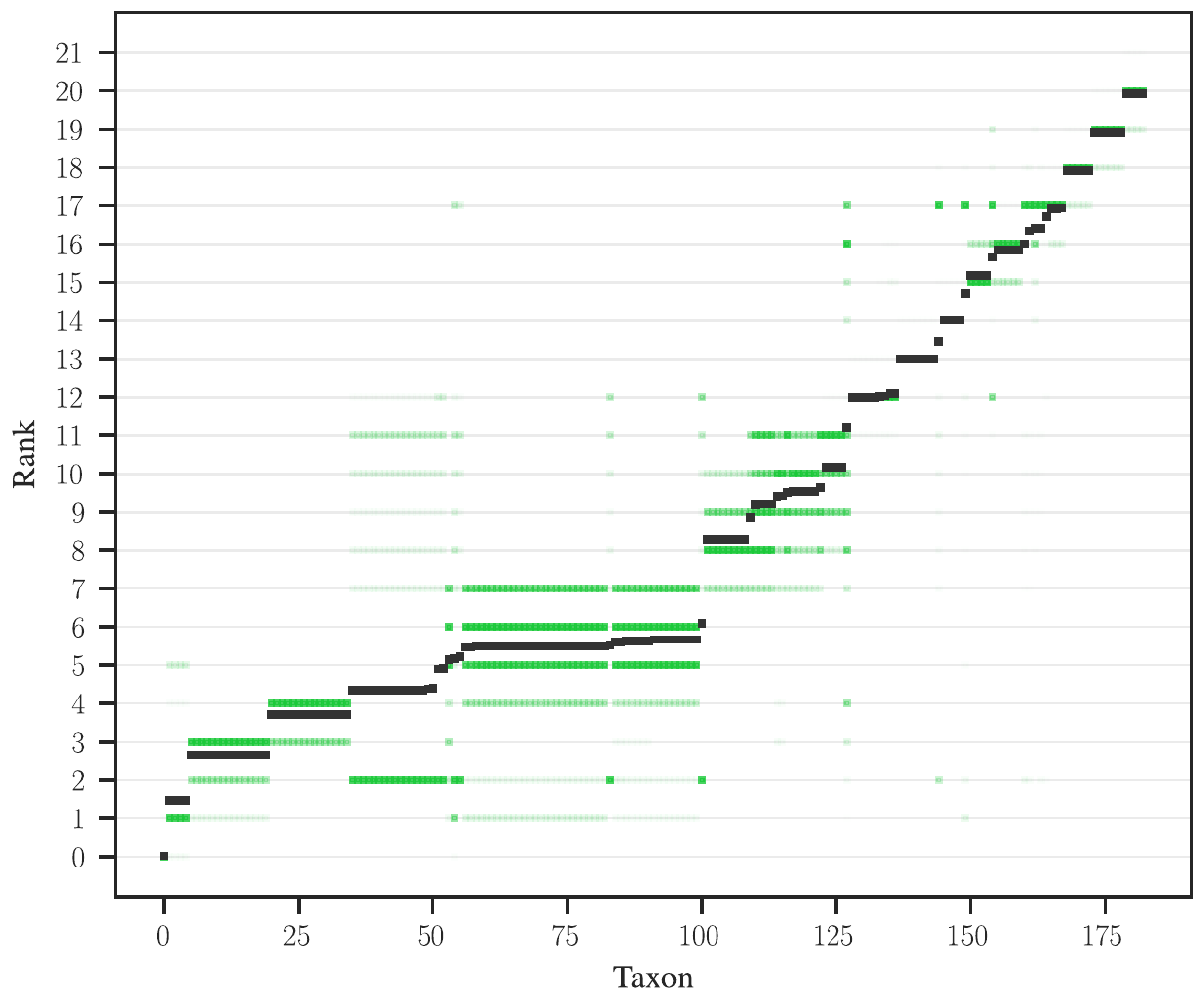}};
      \begin{scope}[
          x={($0.1*(image.south east)$)},
          y={($0.1*(image.north west)$)}]
        \node at (0, 10) {(b)};
      \end{scope}
    \end{tikzpicture}}
  \end{tabular}

  \caption{Inferred ordered group structure of the food web of Little
  Rock lake. Panel (a) shows the identified groups for each taxon, with
  the rank labels shown on the nodes. The edge colors indicate the
  direction: upstream (blue), downstream (red), and lateral
  (grey). Panel (b) shows the marginal posterior distribution of rank
  positions $\pi_i(r)$ (semitransparent green symbols, with opacity
  indicating probability) and mean value $\bar b_i$ (solid black
  symbols), for each species.\label{fig:foodweb}}
\end{figure*}

With this modification of the model, we can perform inference using MCMC
in the same way as with the original model, using only a different
posterior distribution. However, there are some special considerations
that can improve the mixing time when group orderings are relevant,
which we describe in Appendix~\ref{app:mcmc}.

\subsection{Nested SBM}
The uniform prior for the matrix $\bm m$ of Eq.~\ref{eq:prior_m} encodes
the assumption that all matrices are equally likely a priori, and
therefore that the preferences between groups are expected to be
unstructured. Not only is this an unrealistic assumption, but it has
also been shown that it leads to a ``resolution limit,'' where the
maximum number of groups that can be inferred scales as $O(\sqrt{N})$
for sparse networks~\cite{peixoto_parsimonious_2013}. An effective
solution for this problem has been proposed in
Ref.~\cite{peixoto_hierarchical_2014}, where the uniform prior is
replaced by a multigraph SBM, where the nodes are groups and the edge
counts $\bm m$ are the edge multiplicities. The groups and edge counts
of this additional SBM are again modelled as another SBM, forming a
nested hierarchy of SBMs. Since the matrix $\bm m$ is symmetric, we can
replace Eq.~\ref{eq:prior_m} by the undirected prior derived in
Ref.~\cite{peixoto_nonparametric_2017}, which we omit here for brevity
--- the reader can refer to Refs.~\cite{peixoto_hierarchical_2014,peixoto_nonparametric_2017} for a comprehensive description of this modelling approach.

With this modification we can uncover ordered community structures
without such a resolution limit, which is what we will employ in the
rest of this work.

We emphasize that the hierarchical structure present in the nested SBM is of
an entirely different nature than the ordered hierarchies we have been
considering. In the nested model, the hierarchy exists in the
\emph{model structure itself}, i.e. the fact that we have a sequence of
priors and hyperpriors, not necessarily in the actual networks that it
generates.

\section{Preference and ranking}
\label{sec:preference}

We demonstrate how our model can simultaneously accommodate preference
of connections and ranking, by studying the food web of Little Rock
lake~\cite{martinez_artifacts_1991}. In this network the nodes are taxa,
where each taxon is either an individual species, a species subset with
distinct set of predators and preys (e.g. different stages of
development of individuals of the same species), or an aggregate of
similar species. In our representation, a directed edge $i\to j$ exists
if taxon $i$ is eaten by taxon $j$. In Fig.~\ref{fig:foodweb}a we can see
the result of our method applied to this network of $N=183$ nodes. We
can identify $B=22$ ordered taxonomic groups. The vast majority of edges
go upstream, revealing a substantial degree of trophic
ordering---although the network is far from being acyclic, and we can
observe trophic rank violations, cannibalism (self-loops), and
lateral predation within the same trophic group. Overall, the ordering
uncovered matches the trophic structure that is well understood for food
webs of this type: The basal taxon at the bottom of the hierarchy is an
aggregate of microorganisms labelled only ``fine organic matter,'' which
are consumed by a large number of algae species. Intermediary taxa
include insects, crustaceans and fish, whereas taxa at the top of the
hierarchy correspond to decomposers. However, besides the trophic
ordering, we can also identify clear predation preferences that are not
associated directly with rank. For example, taxonomic group $7$ is
predated by group $8$, but not at all by group $9$, which prefers
instead to predate groups $5$ and $1$, predominantly.

Our methodology allows for a more detailed assessment of the group
ordering by inspecting the entire posterior distribution of
Eq.~\ref{eq:posterior}, instead of the single best partition. For
example, we can obtain the marginal rank distribution of node $i$ given
by
\begin{equation}
  \pi_i(r) = \sum_{\bb}\delta_{b_i,r}P(\bb|\A),
\end{equation}
The above mean over all possible partitions $\bb$ sampled from the
posterior distribution can obtained directly from our MCMC sampling
algorithm. (Note that the lack of invariance to label permutation
renders moot issues that complicate the computation of such marginal
probabilities in the case of the unordered
SBM~\cite{peixoto_revealing_2021}.) The above computation allows for a
continuous ranking of the nodes, via the mean
\begin{equation}
  \bar b_i = \sum_r r \pi_i(r),
\end{equation}
and a decoupling of rank and group, in
the sense that nodes that always belong to different groups can in
principle have the same marginal rank distribution. This will happen
when the clustering is due predominantly to preference, and not a
particular position in the hierarchy.

In Fig.~\ref{fig:foodweb}b we show the marginal rank distribution for
the individual taxa, allowing us to identify a fair amount of rank
uncertainty at intermediary levels.

%\FloatBarrier

\section{Degree correction: local vs. global ordering}\label{sec:deg-corr}

We move now to the role of degree correction in our modeling
approach. Typical techniques for ordering nodes in a one-dimensional
hierarchy attempt, in one way or another, to minimize the rank
violations produced by edges that flow in the direction opposite to the
rank relationship. As a result, methods of this kind have the tendency
to produce orderings that are positively correlated with the difference
between out-degree and in-degree of each node,
\begin{equation}
  d_i = k_i^{\text{out}}-k_i^{\text{in}}.
\end{equation}
In other words, a node with high out-degree but low in-degree will tend
to occupy a low position in hierarchy, whereas a node with low
out-degree but high in-degree will tend to occupy a position at the top.

However, we can easily imagine a situation where an arbitrary
out-/in-degree sequence leads to an inherent ordering given by
$d_i$, but the edges of the network are placed otherwise completely
at random. In this scenario, this ordering only conveys information
about the degree sequence itself, not any additional propensity of
placing edges in a manner that respects the ranking of the
nodes. Methods that cannot make this distinction will conflate
out-/in-degree imbalance with a position in the hierarchy that goes
beyond this local property.

Our model allows us to make the distinction between out-/in-degree
imbalance and a more meaningful latent hierarchy because it accepts the
out-/in-degree sequence $\bm k$ as a set of parameters that are largely
independent from the group affinities $\bm e$. In this way, it will put
nodes in different hierarchical levels only if there is sufficient
evidence to justify a preference that goes beyond degree imbalance.

We illustrate this with a simple artificial network model, where all
nodes have the same total degree $k_i^{\text{out}}+k_i^{\text{in}} = k$,
but the imbalance is given by an out-degree sampled from a binomial
distribution with mean $(N-i)/(N-1)$, i.e.
\begin{multline}\label{eq:imbalanced}
  P(k_i^{\text{out}},k_i^{\text{in}}|k)=\delta_{k_i^{\text{in}},k-k_{\text{out}}}\times\\
  {k\choose k_i^{\text{out}}} \left(\frac{N-i}{N-1}\right)^{k_i^{\text{out}}}
  \left(\frac{i-1}{N-1}\right)^{k-k_i^{\text{out}}}.
\end{multline}
Conditioned on a degree sequence sampled in this manner as a hard
constraint,\footnote{Sampling out-/in-degrees from
Eq.\ref{eq:imbalanced} may result in values for which the total sum of
in- and out-degrees are not identical, which makes a half-edge pairing
impossible. If this happens, we resample values for a node chosen
uniformly at random, repeatedly, until a feasible degree sequence is
obtained.} we then generate a pairing between the corresponding
half-edges uniformly at random, and then obtain a final multigraph $\A$.

\begin{figure}
  \begin{tabular}{cc}
    \begin{overpic}[width=.5\columnwidth]{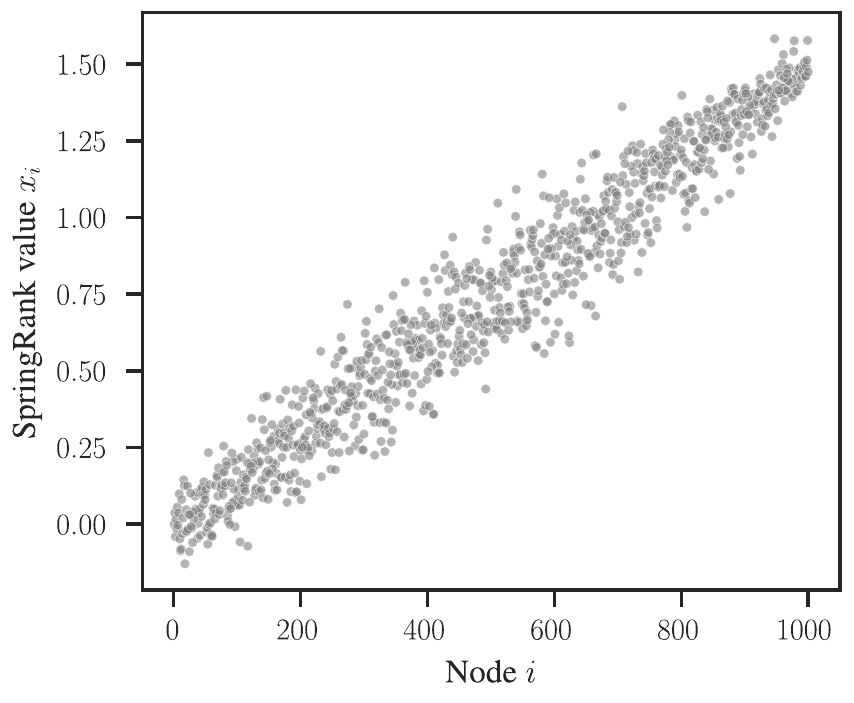}
      \put(0,84){(a)}
    \end{overpic}
    &
    \begin{overpic}[width=.5\columnwidth]{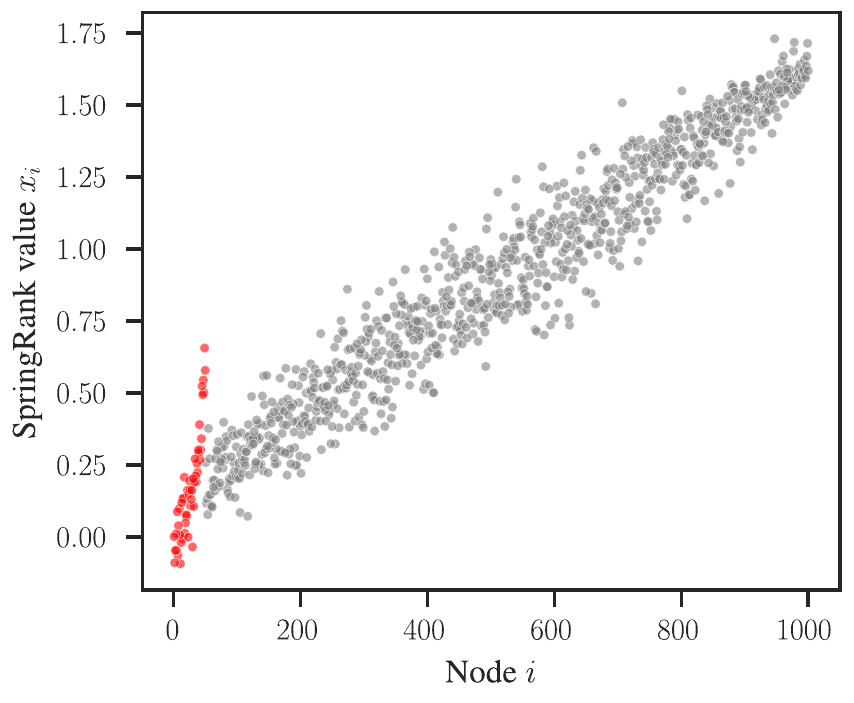}
      \put(0,84){(b)}
    \end{overpic}\\
    \begin{overpic}[width=.5\columnwidth]{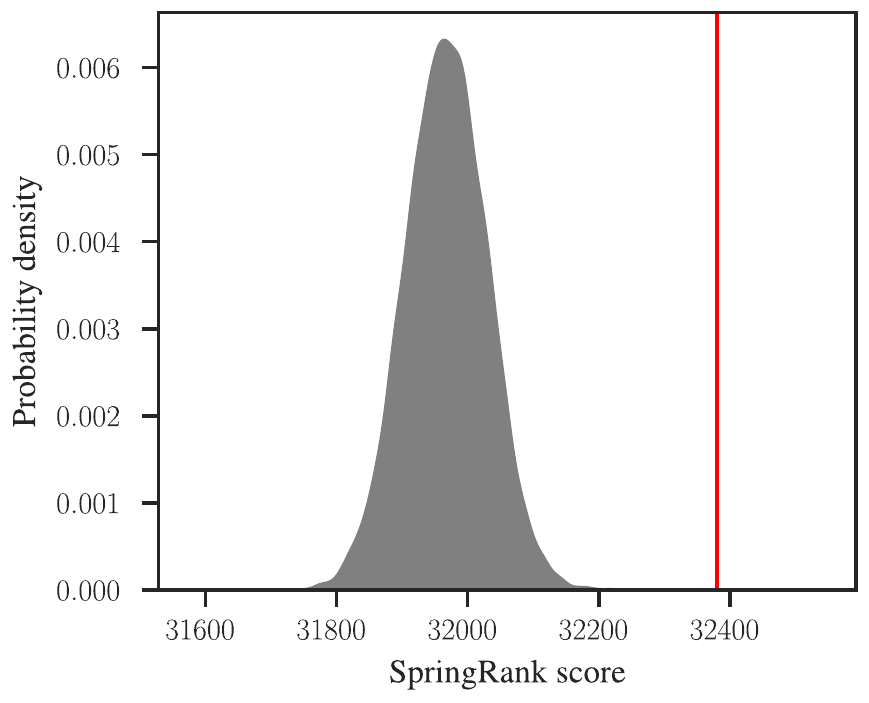}
      \put(0,84){(c)}
    \end{overpic}
    &
    \begin{overpic}[width=.5\columnwidth]{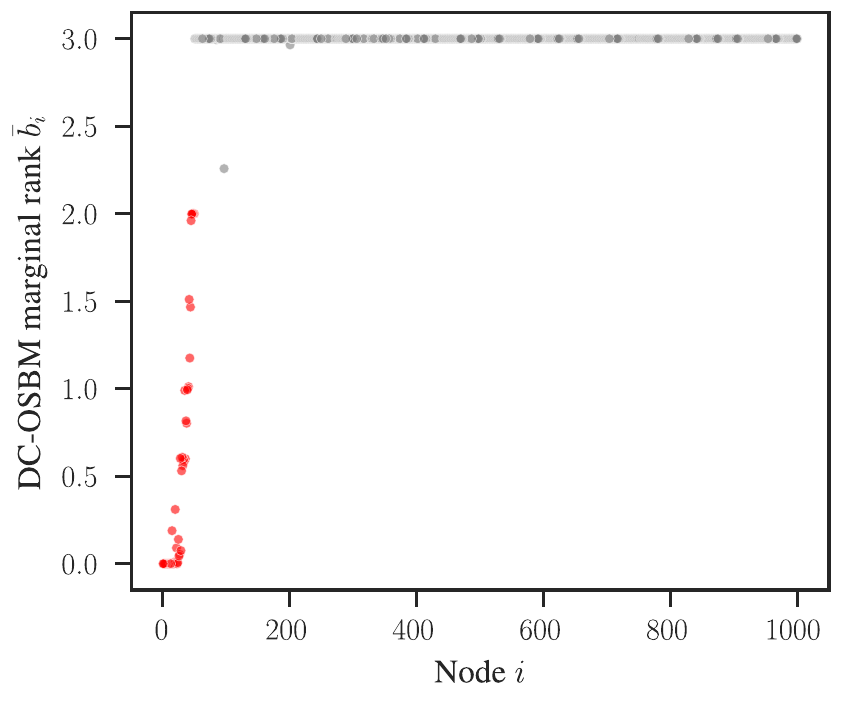}
      \put(0,84){(d)}
    \end{overpic}
  \end{tabular}

  \caption{(a) SpringRank values for a network sampled uniformly at
  random with imposed in/out-degrees themselves sampled from
  Eq.~\ref{eq:imbalanced}, with $k=50$ and $N=1000$. (b) Same as (a),
  but with 500 additional upstream edges added uniformly at random
  between nodes with index in the range $[1,N/20]$ (shown in red). (c)
  Distribution of SpringRank score values for networks sampled uniformly
  at random with imposed degree sequence identical to panel (a). The
  solid vertical line marks the value obtained for the network
  considered in (b). (d) Marginal rank $\bar b_i$ obtained with the
  DC-OSBM for the same network as in panel (b).\label{fig:springrank}}
\end{figure}
\begin{figure}
  \includegraphics[width=\columnwidth]{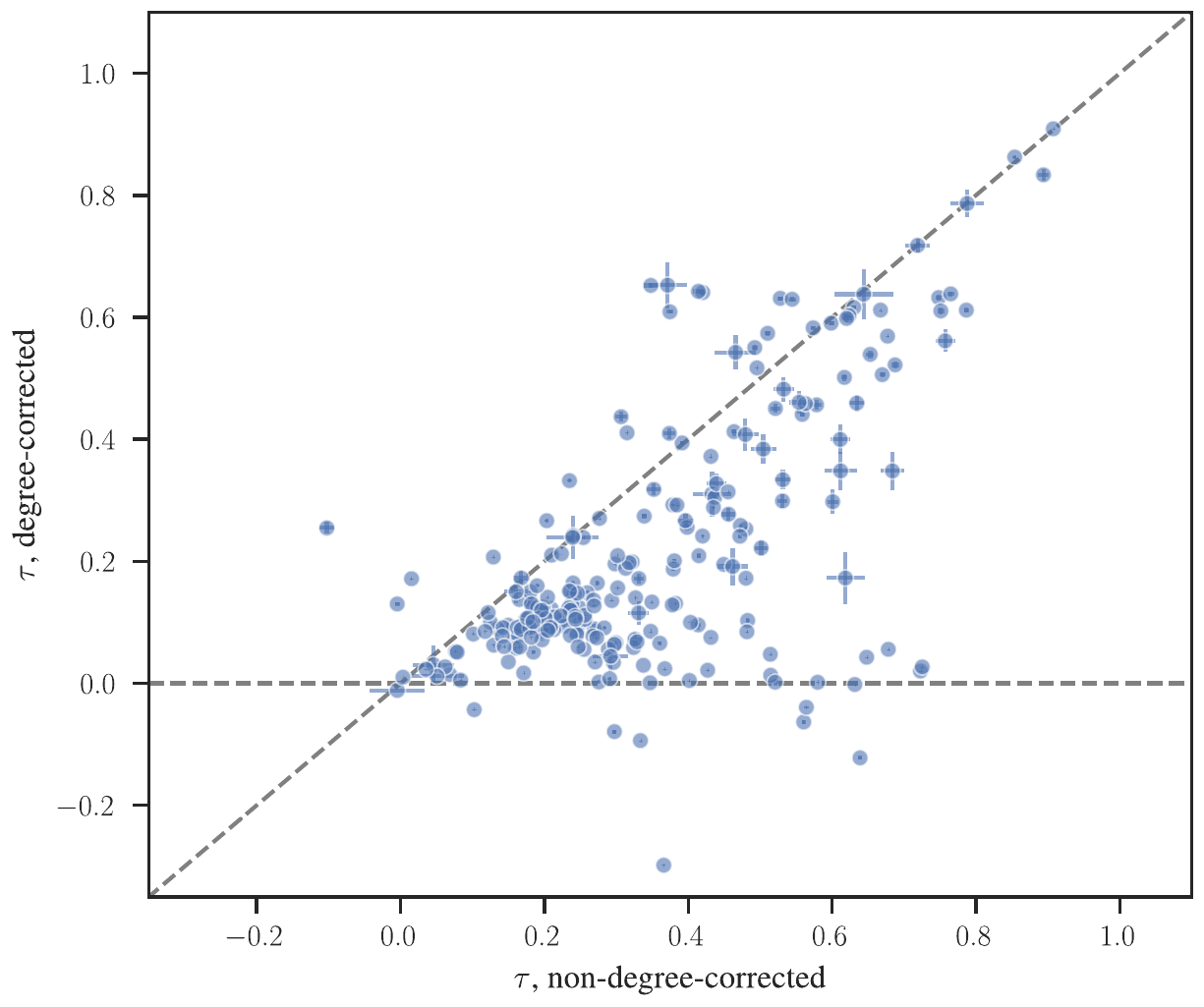}
  \caption{Comparison of Kendall's rank correlation coefficient $\tau$
between the degree imbalance $d_i$ and rank $b_i$ for each network in
our dataset, for both the degree-corrected and non-degree-corrected
version of our model. The sloped dashed line shows the diagonal where
the two values are the same.\label{fig:tau}}
\end{figure}

When applied to a network sampled from this model, our approach assigns
all nodes to a single group---meaning that it (correctly) does not
identify any preference of connections that go beyond the degree
sequence. As a comparison, we show in Fig.~\ref{fig:springrank}a the
result obtained with the SpringRank method~\cite{de_bacco_physical_2018}
on the same example. Since this method does not include
degree-correction, it also reveals only the degree imbalance. As a means
of circumventing the identification of spurious hierarchies of this
kind, the authors of Ref.~\cite{de_bacco_physical_2018} have suggested a
null model test, using the rank score provided by the method itself is a
test statistic. Unfortunately, this approach is overly sensitive to
minor deviations from the null model, as we demonstrate in the
following. After generating a network from the above model, we modify
the sampled network by adding a small number of random upstream edges
involving only the first $5\%$ of the nodes (i.e. nodes with index $1$
to $N/20$). The result, as we can see in Fig.~\ref{fig:springrank}c, is
that the statistical test (correctly) rejects the null model, while the
inferred rankings still predominantly reveal only the degree imbalance
for the majority of the nodes (Fig.~\ref{fig:springrank}b). This is very
much the same problem we encounter when using null model rejection to
prevent the detection of spurious communities when doing community
detection~\cite{peixoto_descriptive_2022}: the statistical significance
of a global quality score tells us very little about the statistical
significance of the actual latent variables uncovered---the questions
``is the value of the quality score significant?'' and ``are the
inferred latent variables significant?'' are not equivalent, and the
answer to the first serves as a very poor proxy to the second.
Ultimately, the rejection of a null model tells us what kind of
structure a network does not have, but cannot tell us what structure it
does have. Because of this problem, with a method such as SpringRank, it
is not in general possible in uncontrolled empirical settings to fully
distinguish between degree imbalance and statistically significant
non-local hierarchies.

Since our approach is based on the inference of a flexible generative
model, rather than the rejection of a null model, we are able to deal
with the above situation in a more satisfying manner. In
Fig.~\ref{fig:springrank}d we show the inferred rankings of same
modified network considered above, according to the degree-corrected
ordered SBM (DC-OSBM). Due to degree-correction, the method puts all
unperturbed nodes into a single hierarchical level---despite their
varied out-/in-degree imbalance---and the perturbed nodes into lower
levels, reflecting the upstream edges that were added between them. The
interpretation becomes more straightforward: the structure of the first
$N/20$ nodes cannot be explained solely by the out-/in-degree imbalance,
and the model reveals instead a non-local ordering.

Degree correction is a property that is optional in our approach. It can
be ``turned off'' by choosing an alternative prior for the degree
sequence, $P(\bm k|\bm e,
\bb)$~\cite{peixoto_nonparametric_2017}. Therefore, in situations where
degree imbalance is expressively desired as a ranking criterion, our
method can still be used. However, even with degree-correction, it is
still possible to use the degree imbalance to ``locally'' order nodes
that otherwise belong to the same rank, simply by using a
lexicographical partial ordering, i.e. $(b_i,d_i) \le
(b_j,d_j)$ if $b_i<b_j$ or $b_i = b_j$ and $d_i \le
d_j$. More importantly, our approach allows for model selection:
given the same network $\A$, we can decide if the degree-corrected model
variant is more compressive or not, by computing its description length,
and therefore if there is more statistical evidence justifying its
description of the data.

In Fig.~\ref{fig:tau} we show a comparison between the degree-corrected
and non-degree-corrected version of our model for 251 empirical directed
networks of different domains (see Appendix~\ref{app:data} for
descriptions). We compute Kendall's rank correlation coefficient $\tau$
between the degree imbalance $d_i$ and the ranking obtained for
each model, for each network in our dataset. The typical case is that
the correlation with degree imbalance decreases when degree-correction
is used, often substantially, indicating that in those cases the degree
sequence is a major contribution to the inferred hierarchy obtained
without degree-correction, and there is otherwise no significant support
for it. There are also situations when the same correlation
values---sometimes also high---are observed for both model
variants. This indicates that although the degree sequence itself ends
up being informative of the latent hierarchy, this turns out also to be
corroborated by an additional alignment with the group ordering that goes
significantly beyond the degree imbalance. We can also observe a
minority of situations where the correlation increases when
degree-correction is employed, but these are mostly due to artefacts
caused by the number of hierarchical levels changing significantly from
one model to the other.

\section{Model selection: Is there a hierarchy?}
\label{sec:model-selection}

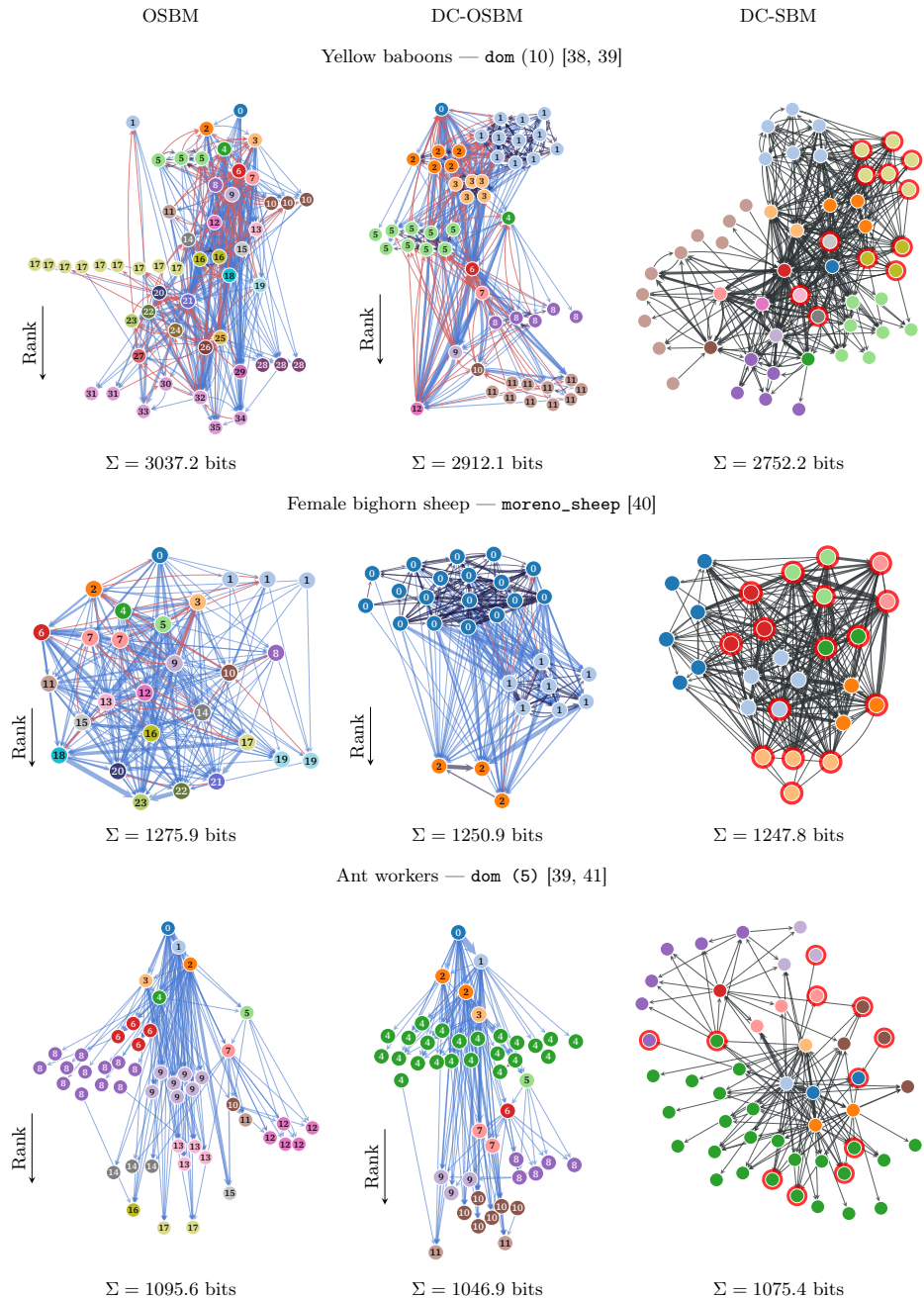
\begin{figure*}
  \resizebox{!}{.37\textheight}{
        \begin{tabular}{ccc}
    OSBM & DC-OSBM & DC-SBM\\[1em]
    \multicolumn{3}{c}{Yellow baboons --- \href{https://networks.skewed.de/net/dom\#Franz_2015e}{\texttt{dom} (10)}~\cite{franz_self-organizing_2015,strauss_domarchive:_2022}}\\
    \raisebox{-\height}{\begin{tikzpicture}
      \node [above right, inner sep=0] (image) at (0,0) {\includegraphics[width=.3\textwidth]{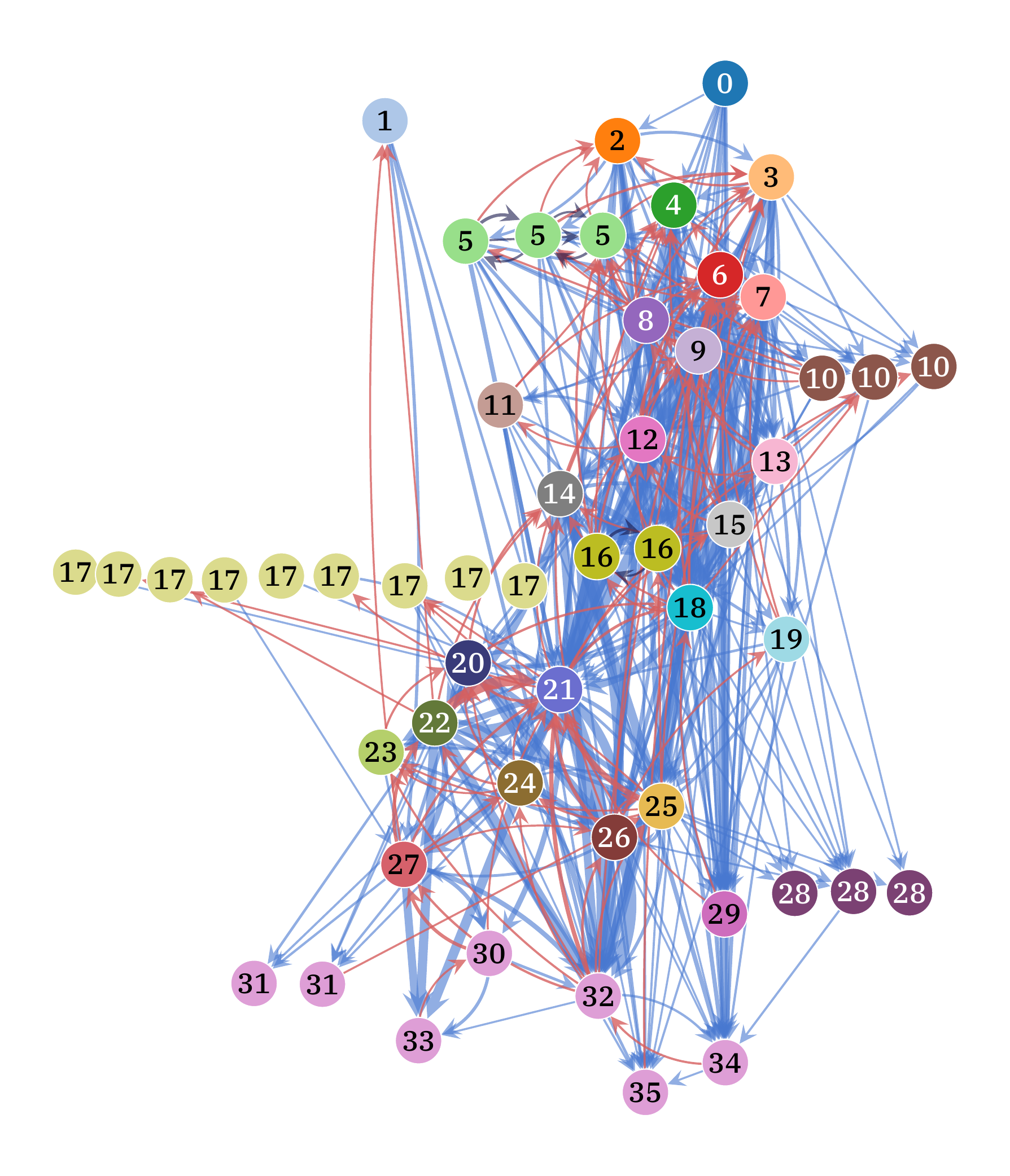}};
      \begin{scope}[
          x={($0.1*(image.south east)$)},
          y={($0.1*(image.north west)$)}]
        \draw [-stealth](1,4) --  node [above,rotate=90] {Rank} (1,2);
      \end{scope}
    \end{tikzpicture}} &
    \raisebox{-\height}{\begin{tikzpicture}
      \node [above right, inner sep=0] (image) at (0,0) {\includegraphics[width=.23\textwidth]{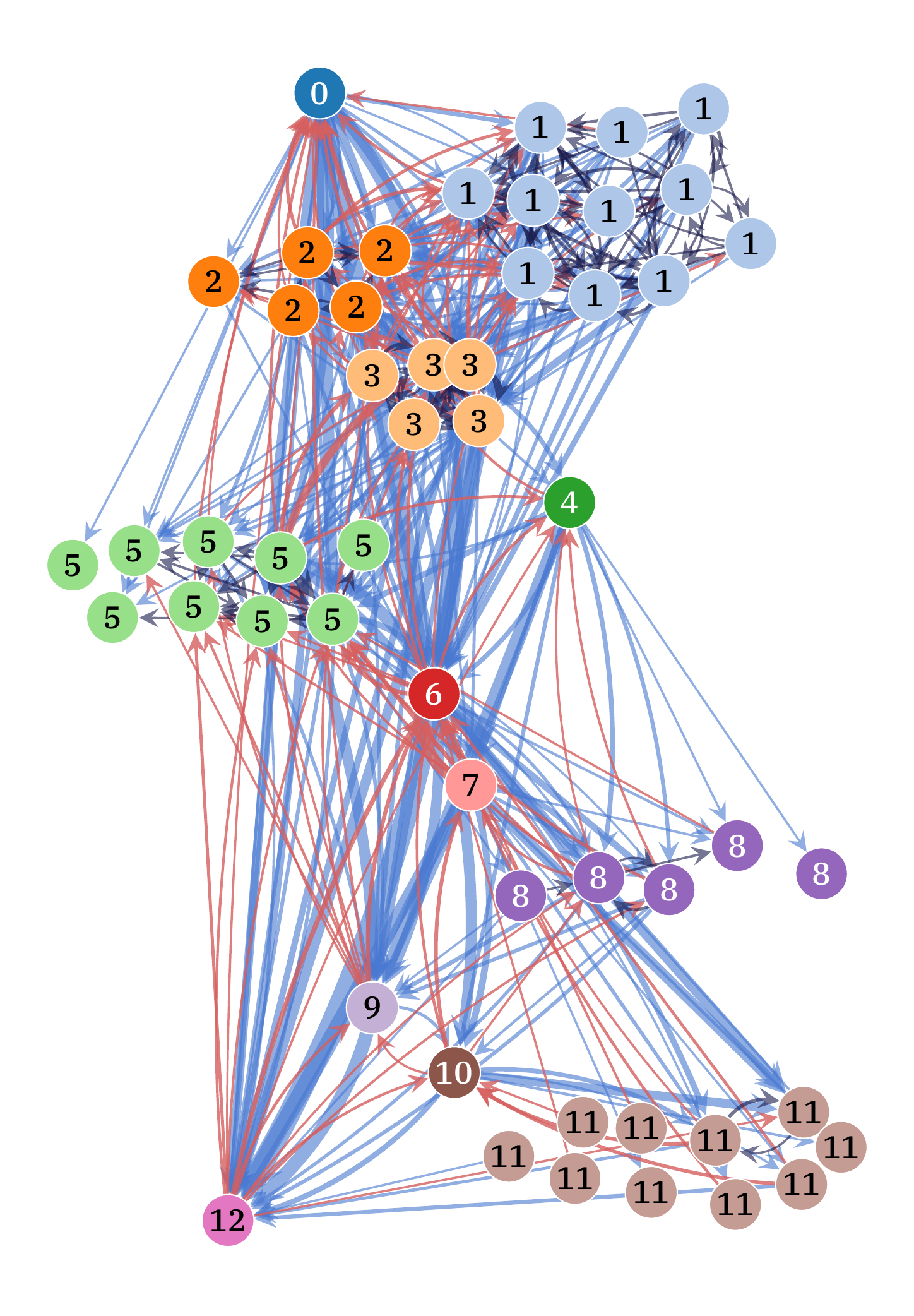}};
      \begin{scope}[
          x={($0.1*(image.south east)$)},
          y={($0.1*(image.north west)$)}]
        \draw [-stealth](1,4) --  node [above,rotate=90] {Rank} (1,2);
      \end{scope}
    \end{tikzpicture}} &
    \raisebox{-\height}{\begin{tikzpicture}
      \node [above right, inner sep=0] (image) at (0,0) {\includegraphics[width=.3\textwidth]{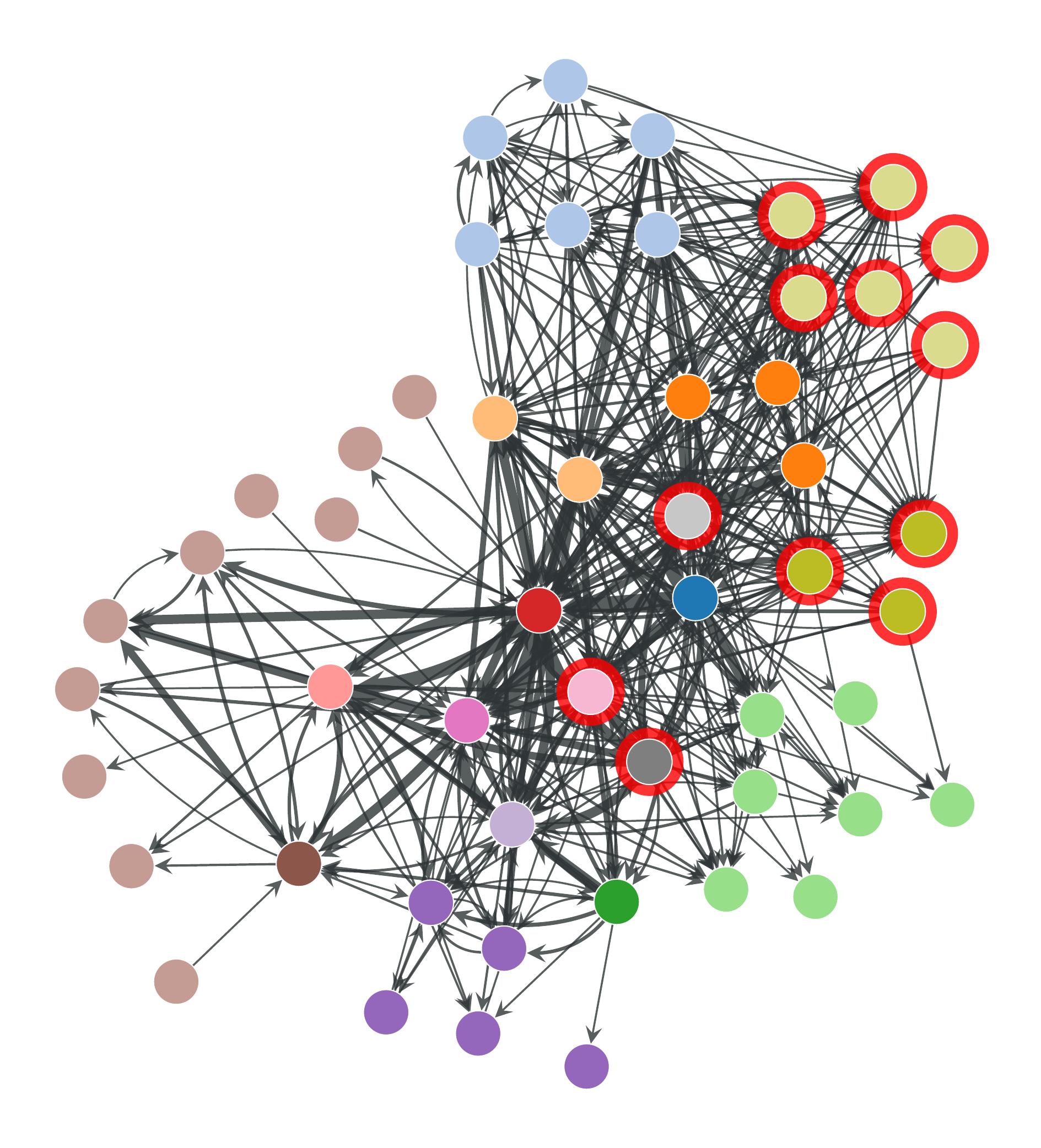}};
      \begin{scope}[
          x={($0.1*(image.south east)$)},
          y={($0.1*(image.north west)$)}]
      \end{scope}
    \end{tikzpicture}}\\
    $\Sigma = 3037.2$ bits &
    $\Sigma = 2912.1$ bits &
    $\Sigma = 2752.2$ bits \\[1em]
    \multicolumn{3}{c}{Female bighorn sheep --- \href{https://networks.skewed.de/net/moreno_sheep}{\texttt{moreno\_sheep}}~\cite{hass_social_1991}}\\
    \raisebox{-\height}{\begin{tikzpicture}
      \node [above right, inner sep=0] (image) at (0,0) {\includegraphics[width=.3\textwidth]{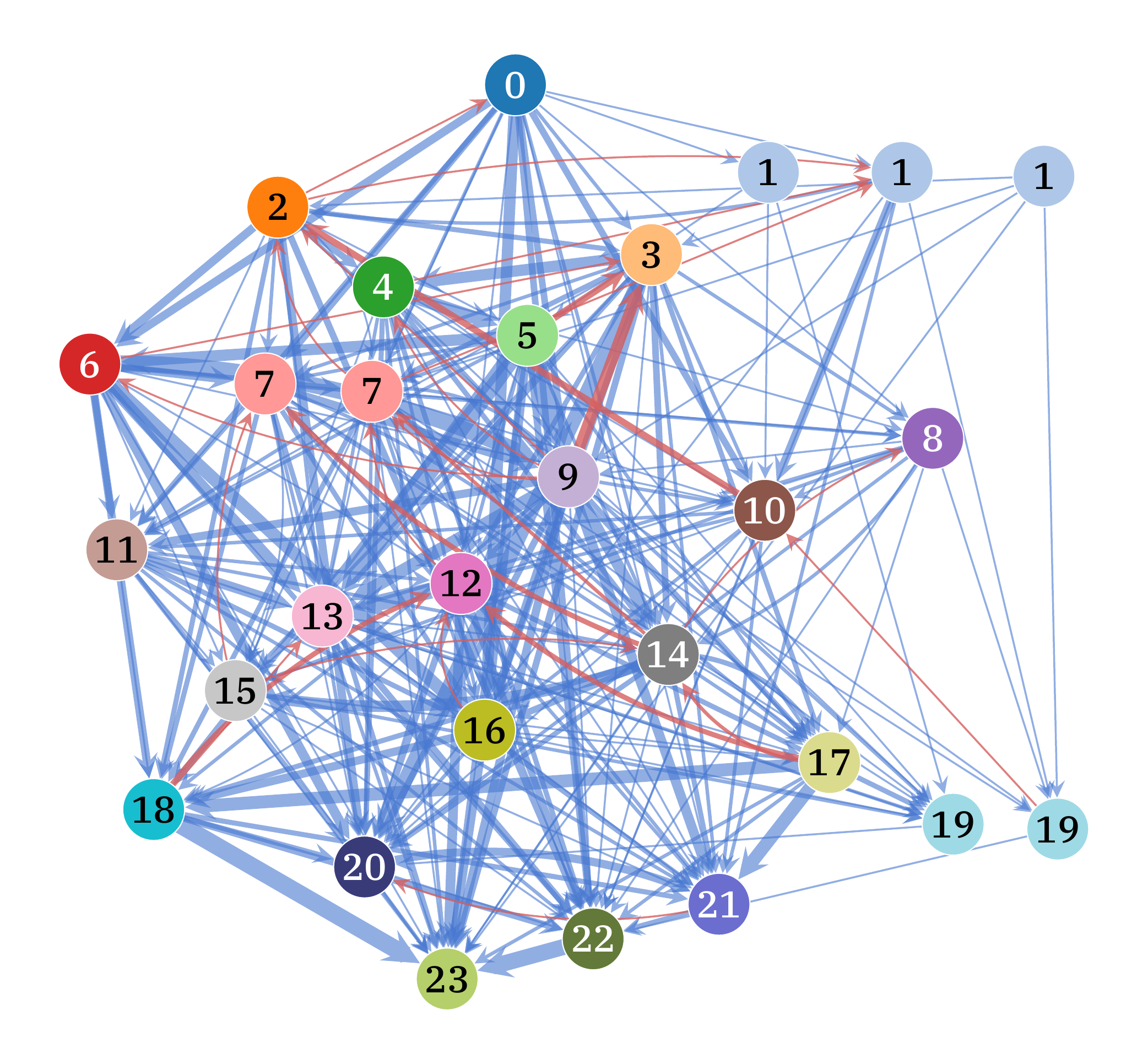}};
      \begin{scope}[
          x={($0.1*(image.south east)$)},
          y={($0.1*(image.north west)$)}]
        \draw [-stealth](.5,4) --  node [above,rotate=90] {Rank} (.5,2);
      \end{scope}
    \end{tikzpicture}} &
    \raisebox{-\height}{\begin{tikzpicture}
      \node [above right, inner sep=0] (image) at (0,0) {\includegraphics[width=.25\textwidth]{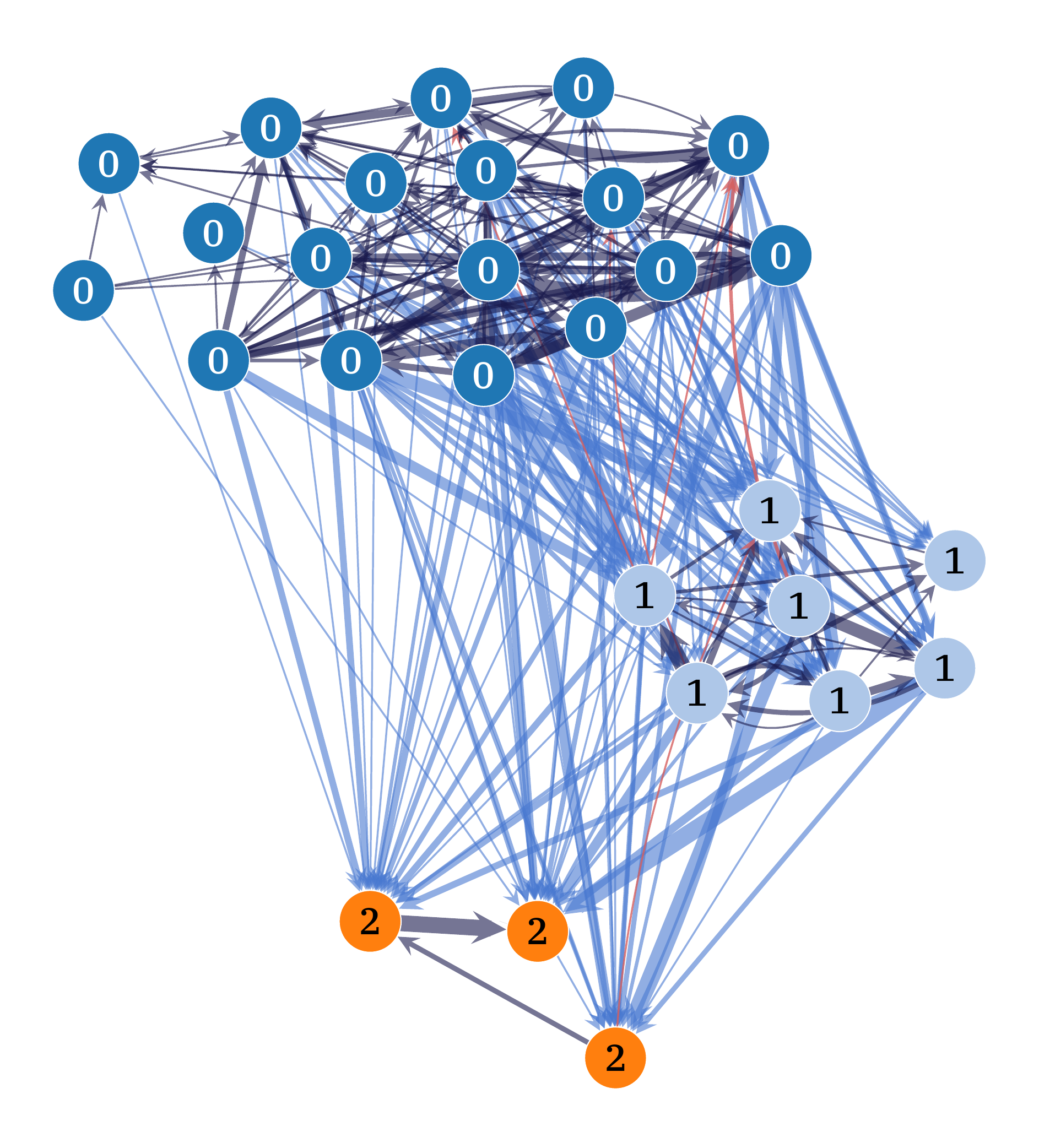}};
      \begin{scope}[
          x={($0.1*(image.south east)$)},
          y={($0.1*(image.north west)$)}]
        \draw [-stealth](1,4) --  node [above,rotate=90] {Rank} (1,2);
      \end{scope}
    \end{tikzpicture}} &
    \raisebox{-\height}{\begin{tikzpicture}
      \node [above right, inner sep=0] (image) at (0,0) {\includegraphics[width=.25\textwidth]{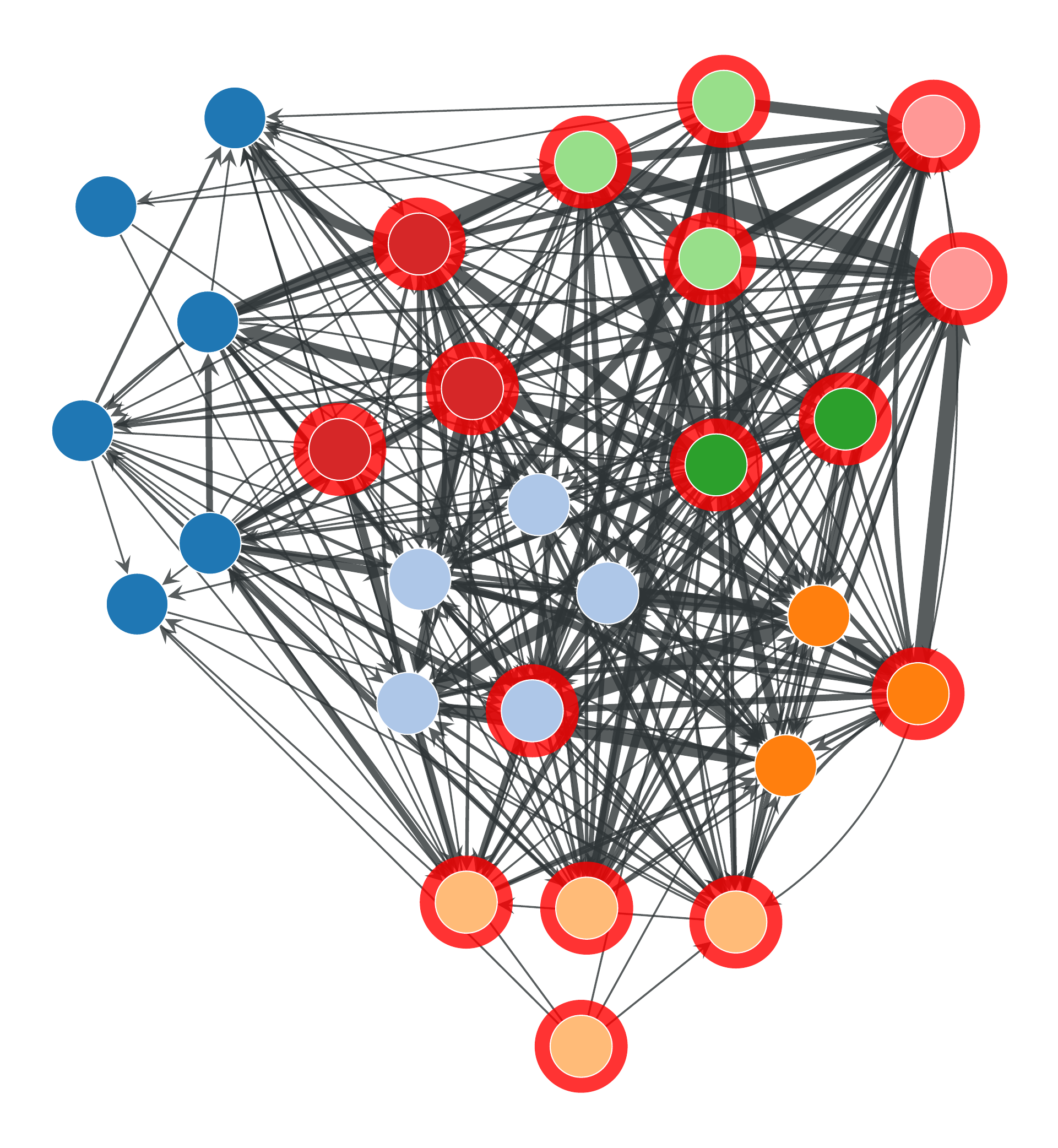}};
      \begin{scope}[
          x={($0.1*(image.south east)$)},
          y={($0.1*(image.north west)$)}]
      \end{scope}
    \end{tikzpicture}}\\
    $\Sigma = 1275.9$ bits &
    $\Sigma = 1250.9$ bits &
    $\Sigma = 1247.8$ bits\\[1em]
    \multicolumn{3}{c}{Ant workers --- \href{https://networks.skewed.de/net/dom\#Shimoji_2014c}{\texttt{dom (5)}}~\cite{shimoji_global_2014,strauss_domarchive:_2022}}\\
    \raisebox{-\height}{\begin{tikzpicture}
      \node [above right, inner sep=0] (image) at (0,0) {\includegraphics[width=.3\textwidth]{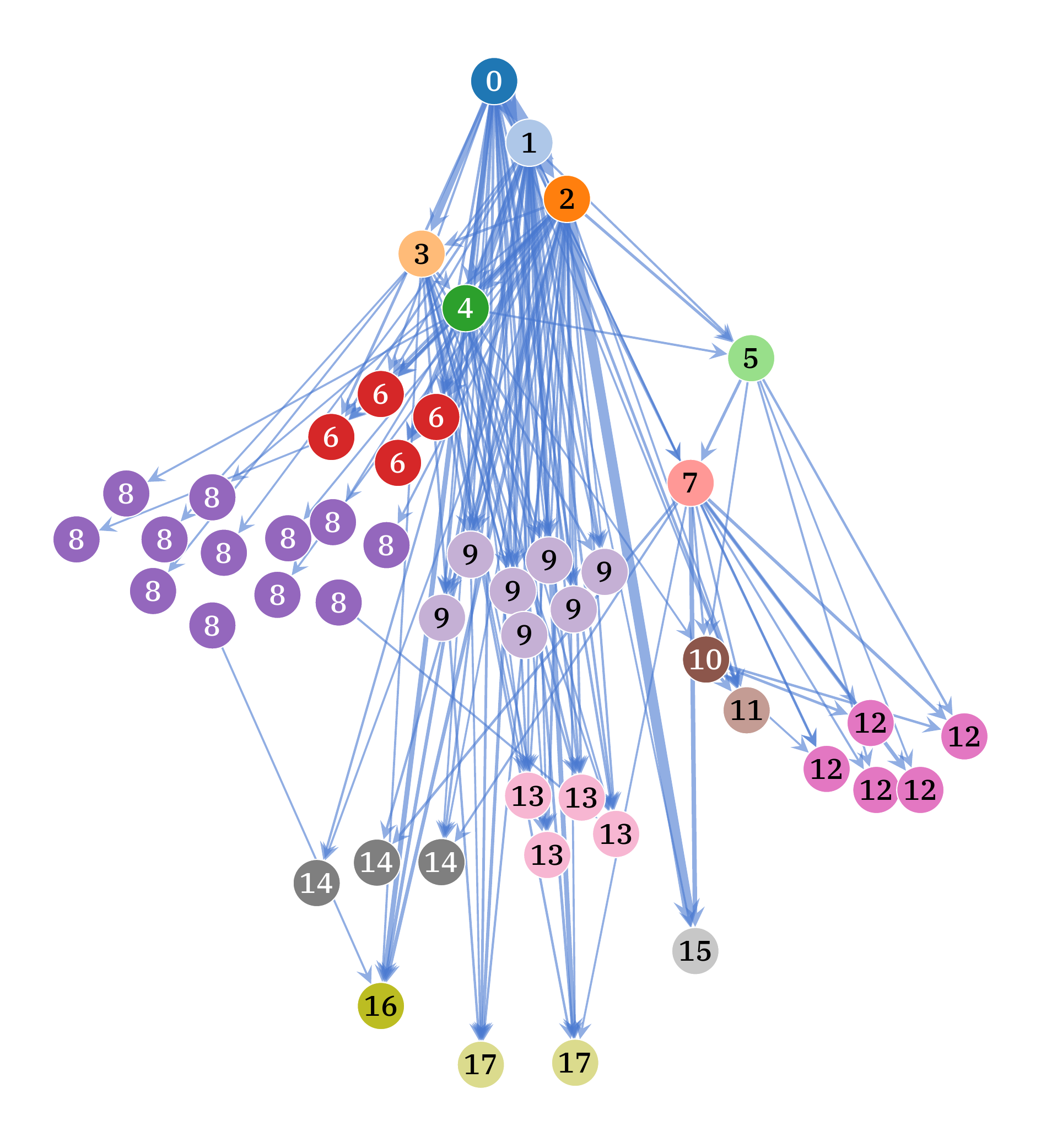}};
      \begin{scope}[
          x={($0.1*(image.south east)$)},
          y={($0.1*(image.north west)$)}]
        \draw [-stealth](.5,4) --  node [above,rotate=90] {Rank} (.5,2);
      \end{scope}
    \end{tikzpicture}} &
    \raisebox{-\height}{\begin{tikzpicture}
      \node [above right, inner sep=0] (image) at (0,0) {\includegraphics[width=.22\textwidth]{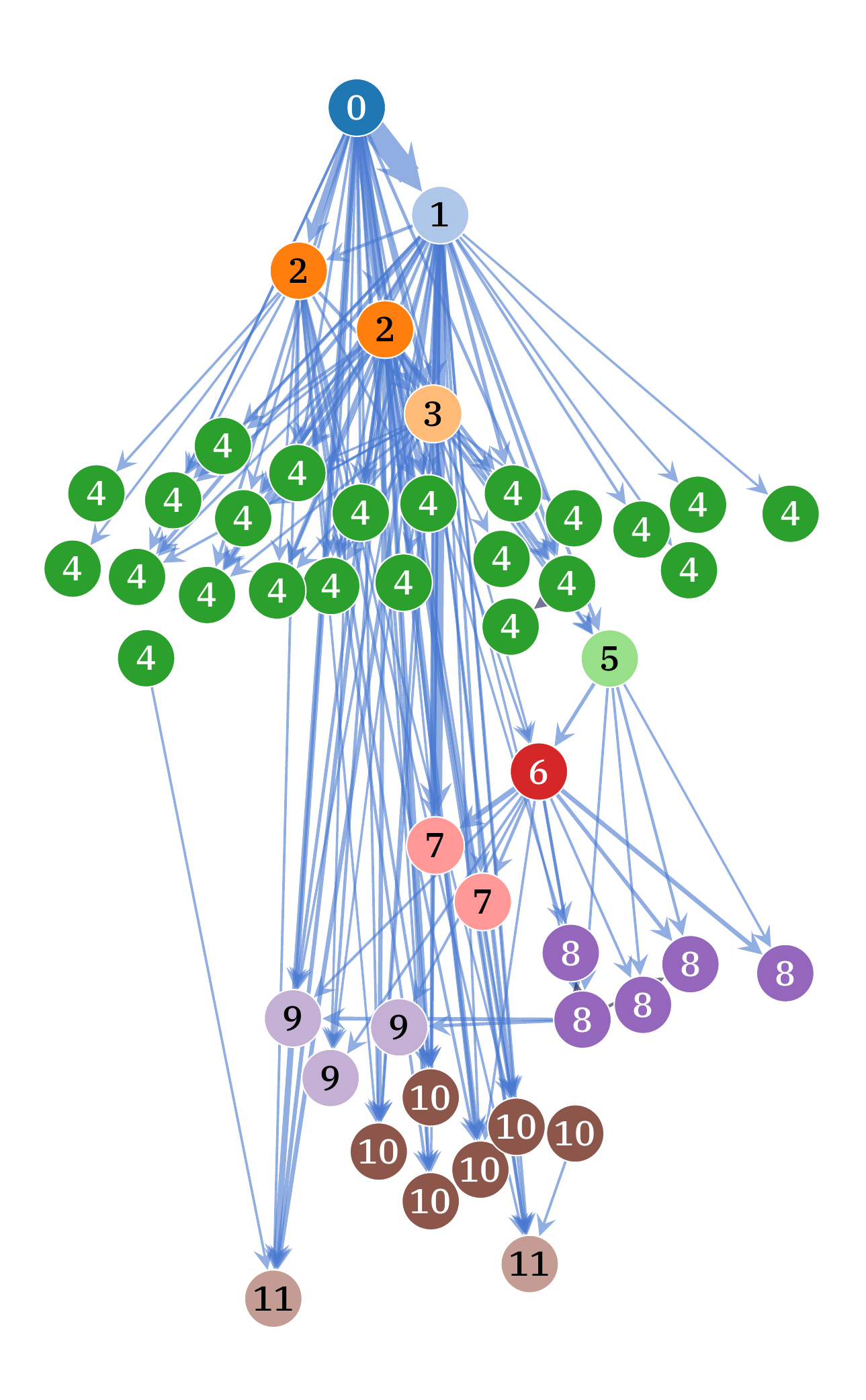}};
      \begin{scope}[
          x={($0.1*(image.south east)$)},
          y={($0.1*(image.north west)$)}]
        \draw [-stealth](1,4) --  node [above,rotate=90] {Rank} (1,2);
      \end{scope}
    \end{tikzpicture}} &
    \raisebox{-\height}{\begin{tikzpicture}
      \node [above right, inner sep=0] (image) at (0,0) {\includegraphics[width=.3\textwidth]{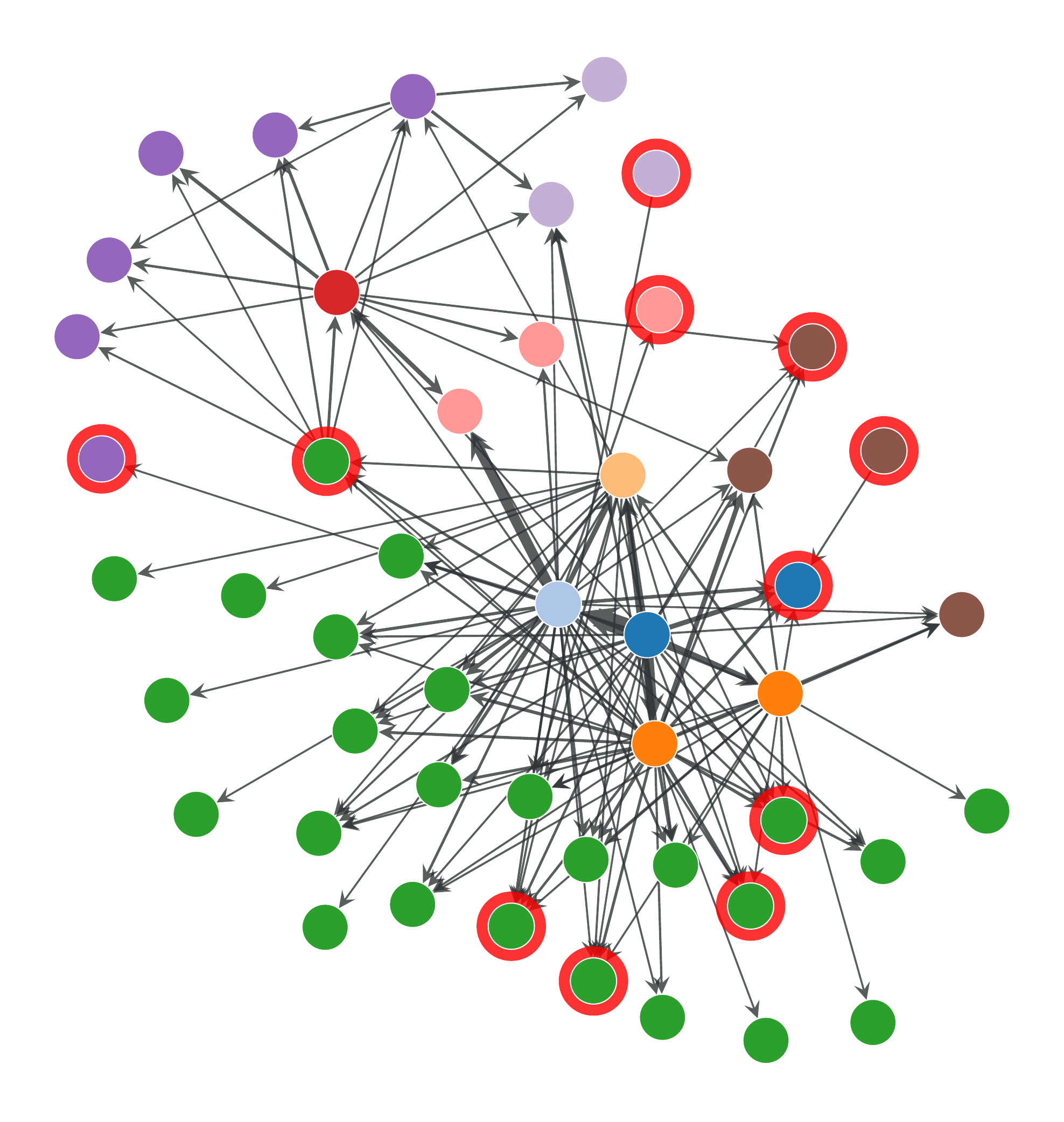}};
      \begin{scope}[
          x={($0.1*(image.south east)$)},
          y={($0.1*(image.north west)$)}]
      \end{scope}
    \end{tikzpicture}}\\
    $\Sigma = 1095.6$ bits &
    $\Sigma = 1046.9$ bits &
    $\Sigma = 1075.4$ bits
  \end{tabular}}

  \caption{Inferred dominance hierarchy and community
  structure of antagonistic animal behavior. The columns from left to
  right contain the results of the non-degree-corrected ordered SBM
  (OSBM), the degree-corrected ordered SBM (DC-OSBM), and the
  degree-corrected SBM (DC-SBM). The rows, from top to bottom, show the
  antagonistic interactions for a group of yellow
  baboons~\cite{franz_self-organizing_2015}, female bighorn
  sheep~\cite{hass_social_1991}, and ant
  workers~\cite{shimoji_global_2014}.  Each panel shows the identified
  groups for each individual, with the rank labels shown on the
  nodes---except for the rightmost column, where the groups are not
  ordered. For the first two leftmost columns, the edge colors indicate
  the direction: upstream (blue), downstream (red), and lateral (grey).
  The colors for the rightmost colum match the maximum matching with the
  middle column, and with the unmatched nodes highlighted in red. The
  panels show also the description length value for each fit.
  \label{fig:dom}}
\end{figure*}

\begin{figure*}
  \includegraphics[width=\textwidth]{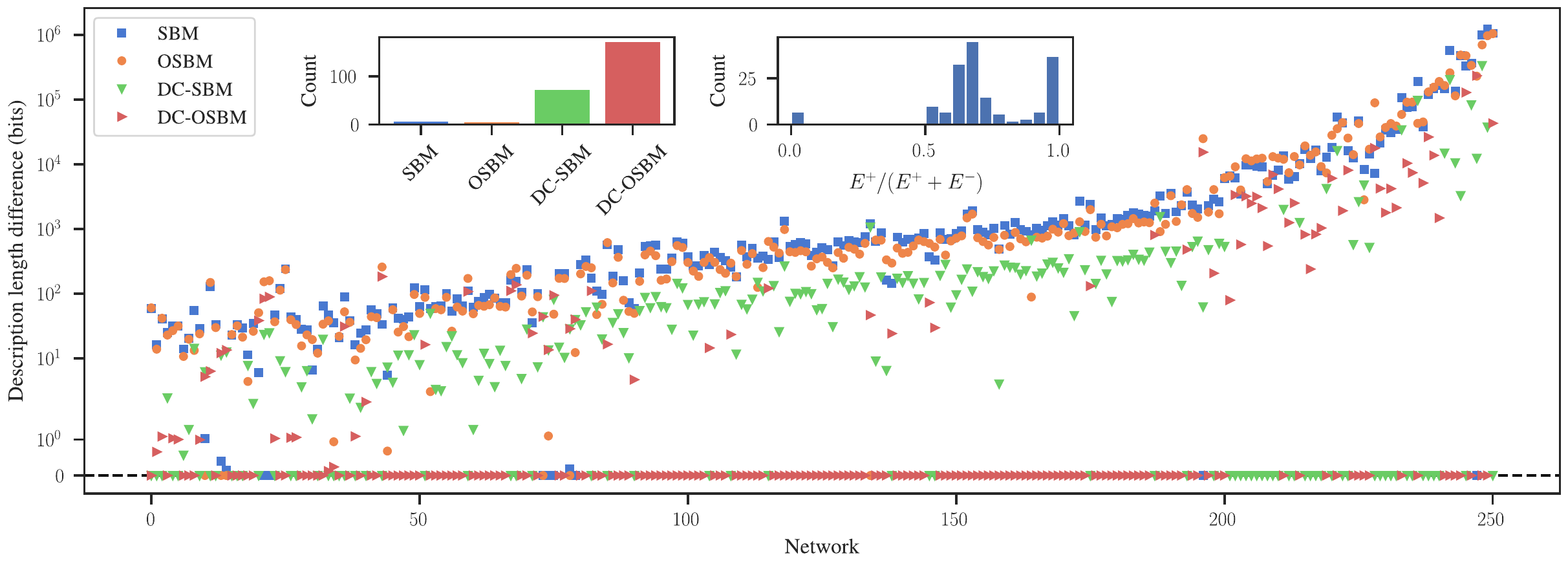} \caption{Comparison
  between models for 251 empirical directed networks, listed in
  Appendix~\ref{app:data}. The values shown are the description length
  differences with respect to the best model, as indicated in the
  legend. The networks are ordered by the minimum description length
  value. The left inset shows the counts that each model type yields a
  shorter description length, and the right inset shows the distribution
  of fraction of upstream edges [$E^+/(E^++E^-)$ or zero if $E=E^0$] for
  networks that are best modelled by the DC-OSBM.\label{fig:empirical}}
\end{figure*}

Given an arbitrary directed network, it is often possible to order its
nodes in such a way that the majority of edges ends up following a
preferred direction according to that ordering. However, by itself,
finding such an ordering is not evidence that it in fact had any role in
the formation of the network---in the same manner that finding
assortative communities in maximally random
networks~\cite{guimera_modularity_2004} is not informative of its
generative process~\cite{peixoto_descriptive_2022}.

A tempting approach to evaluate the statistical significance of a node
ordering is to compare it with what can be obtained with a null model,
e.g. a network with the same out-/in-degree sequence, but otherwise
sampled uniformly at random. This is more easily done via a proxy scalar
statistic, such as the total number of rank violations. But as we have
already seen in the previous section, this approach, although
straightforward, can be quite misleading, since the significance of such
global quantities can be very poorly informative of the significance of
the actual rankings observed. As seen in Fig.~\ref{fig:springrank} we
can obtain overall ``significant'' results by manipulating only a small
minority of the edges of the network. It is important to emphasize that
this is not simply a technical problem that can be circumvented by
tweaking the test statistic; instead it is a fundamental limitation of
null model testing, which is only capable of answering the following
question with ``yes'' or ``no'': can the null model be rejected with
some confidence? A ``no'' answer does not give any information about how
the null model is likely to be true, and a ``yes'' answer can tell us
nothing more than how the network was \emph{not} generated---no further
details of its generative process can be inferred from this test,
including any ranking of its nodes.

A more robust alternative to the rejection of null models is model
selection: we articulate a variety of generative models as alternative
hypotheses, and check which one is more supported by the data. For the
particular problem at hand, we can compare alternative versions of the
SBM, containing any combination of degree-correction and latent
ordering, in how well they can describe the data. Given the same network
$\A$ and two model choices $\mathcal{H}_1$ and $\mathcal{H}_2$, and
their uncovered partitions $\bb^{(1)}$ and $\bb^{(2)}$, respectively,
this comparison is done via the posterior odds ratio,
\begin{align}
  \Lambda &= \frac{P(\mathcal{H}_1,\bb^{(1)}|\A)}{P(\mathcal{H}_2,\bb^{(2)}|\A)}
  = \frac{P(\bb^{(1)},\A|\mathcal{H}_1)P(\mathcal{H}_1)}{P(\bb^{(2)},\A|\mathcal{H}_2)P(\mathcal{H}_2)}\\
  &= \frac{P(\mathcal{H}_1)}{P(\mathcal{H}_2)} 2^{\Sigma_{\mathcal{H}_2}(\A,\bb^{(2)})-\Sigma_{\mathcal{H}_1}(\A,\bb^{(1)})},
\end{align}
with
$\Sigma_{\mathcal{H}_i}(\A,\bb^{(i)})=-\log_2P(\bb^{(i)},\A|\mathcal{H}_i)$
being the description length of the data according to model
$\mathcal{H}_i$ and its partition $\bb^{(i)}$. Therefore, if we are a
priori agnostic with $P(\mathcal{H}_1) = P(\mathcal{H}_2)$, we should a
posteriori select the model with the shortest description length, and
the difference between them will give us the confidence in our
selection.

As a case study of the application of the above methodology, we turn to
networks of antagonistic behavior between
animals~\cite{strauss_centennial_2022}. A directed antagonistic
relationship between two animals $j\to i$ means that individual $j$
prevails after an aggressive encounter with individual $i$. The overall
dominance of $j$ over $i$ is recorded in the multigraph adjacency matrix
$A_{ij}$ as the number of times this particular outcome was
observed. Such antagonistic relationships are assumed to reveal a
dominance hierarchy in animal societies, the position in which is
believed to influence an individual's access to resources, its chance of
survival and reproduction~\cite{strauss_centennial_2022}.

In Fig.~\ref{fig:dom} we show the results of some model variants for
antagonistic networks of yellow baboons, female bighorn sheep, and ant
workers. We consider the non-degree-corrected ordered SBM (OSBM), the
degree-corrected ordered SBM (DC-OSBM), and the degree-corrected
unordered SBM (DC-SBM). In all cases, the degree-corrected variants
yield a shorter description length, indicating that out-/in-degree
variability can be largely decoupled from mesoscale mixing
patterns. Between the ordered models, the degree-corrected variant
yields a smaller number of groups, with a clearer hierarchical
structure. However, when compared to the unordered model, the results
are mixed. For the yellow baboons, the unordered model yields a
significantly improved compression, meaning that heterogeneity of
preference and direction of interactions is not optimally captured by
the ordered model. This indicates that, although clear asymmetries of
outcomes do exist, they cannot be convincingly ascribed to a
one-dimensional ordering, even if it simultaneously accounts for
group-level preferences. The model variant that discards the inherent
ordering can in this case find a more parsimonious description of this
network, even though it finds a partition that largely (but not
completely) agrees with the ordered model. The results for female
bighorn sheep are similar, but far less conclusive: the difference
between the description length values from the DC-OSBM and DC-SBM is
quite small, yielding only an insignificant posterior odds ratio of
$\Lambda\approx 8.6$ in favor of the unordered model. In such a
situation we cannot reliably evaluate if the lack of evidence for
hierarchy is significant, specially since the partitions yielded by both
models differ substantially, and therefore we must conclude that both
models offer competing but approximately equally plausible accounts of
the data. Finally, the results for the ant worker interactions point in
the other direction, and indicate that the ordered model offers a more
parsimonious description---indeed in this case the network is completely
acyclic, and the inferred model contains only upstream edges.

\begin{figure}
  \includegraphics[width=\columnwidth]{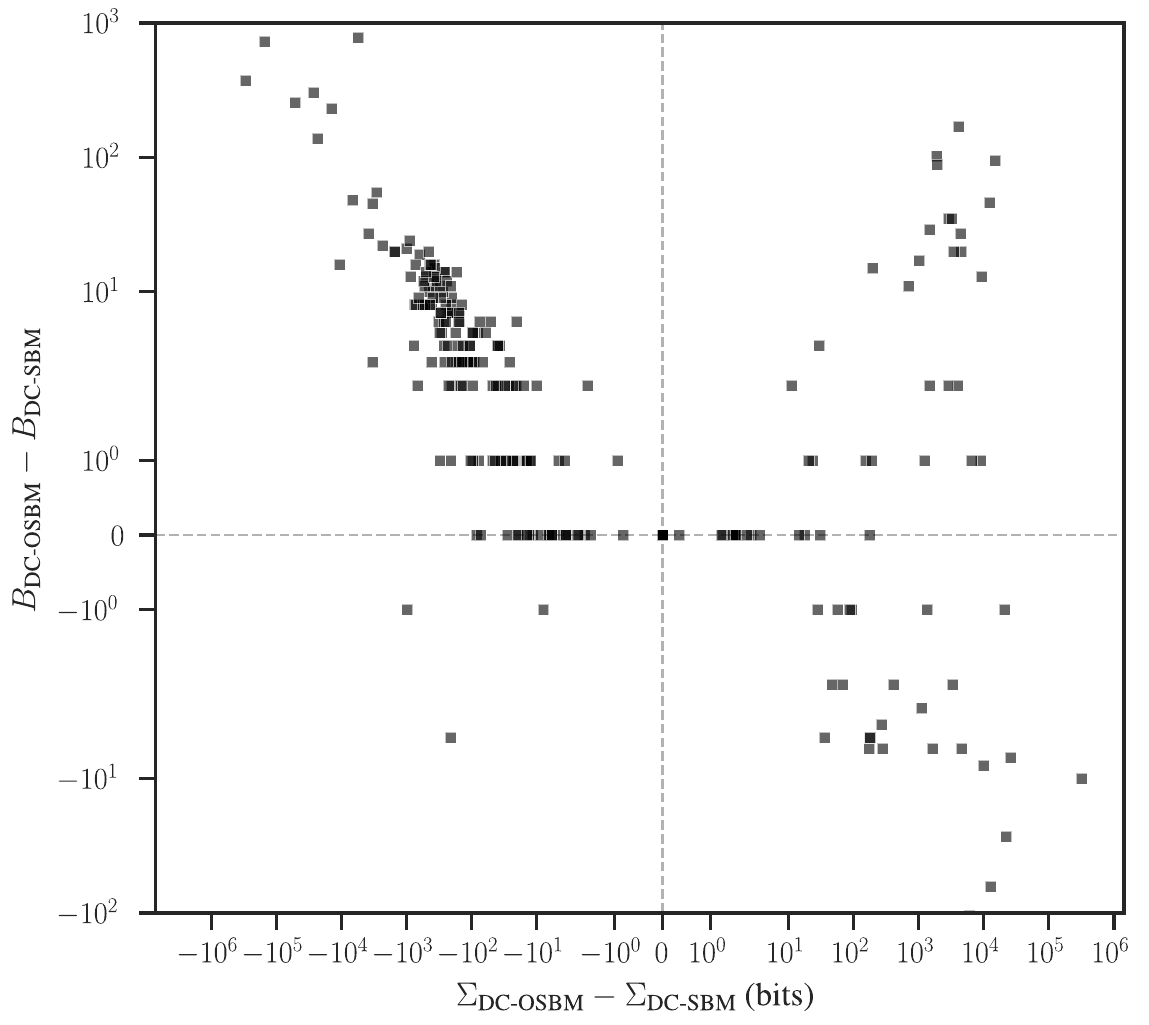}

  \caption{Comparison between the difference in description lengths
  between the DC-OSBM ($\Sigma_{\text{DC-OSBM}}$) and DC-SBM
  ($\Sigma_{\text{DC-SBM}}$) and their respective difference in number
  of groups, $B_{\text{DC-OSBM}}$ and $B_{\text{DC-SBM}}$, obtained for
  the networks in Fig.~\ref{fig:empirical} (each point corresponds to an
  individual network).\label{fig:dBdS}}
\end{figure}

As the examples above show, the most compressive network representations
do not necessarily incorporate rankings between the nodes, although in
all cases we can find such an ordering that initially may seem
plausible. In Fig.~\ref{fig:empirical} we show a more comprehensive
comparison between the ordered and unordered SBMs for a wider set of 251
empirical networks, from diverse domains, listed in
Appendix~\ref{app:data}. For this dataset we find that in fact the
DC-OSBM happens to be the most compressive model for a majority of them,
with the DC-SBM in the second place. Therefore, it does seem to be the
case that node ordering provides opportunities for compression for many
of the networks considered, although the several exceptions mean that
ultimately this needs to be evaluated in a case-by-case basis. It is
worth observing that even when the ordered model is selected, as we
discussed previously, this does not necessarily mean that the rank
alignment is large; this could simply be due to an overall rank
coherence. Indeed, as we can see in the right inset of
Fig.~\ref{fig:empirical}, the rank alignment distribution is bimodal,
with an abundance of networks with moderate values, and another group
with very high values, and hence a more prominent hierarchical
structure.

It is useful to remark on the possibly counter-intuitive fact that the
ordered versions of the SBM can exploit rank coherence for compression,
even when rank alignment is minimal, as we had shown in
Fig.~\ref{fig:prior_comp}. This means that in a situation where no
actual alignment exists between the group ordering and edge direction, a
maximal rank coherence will correspond to a full reciprocity of the edge
counts, $e_{rs}=e_{sr}$, which is a special case of the ordered SBMs,
but would occur only with a very small probability according to the
unordered prior, which expects instead asymmetric matrices. As a result,
the ordered SBMs will be selected as the preferred model when a
substantial reciprocity between groups exists, which accounts for many
cases in Fig.~\ref{fig:empirical}. Furthermore, we point out that since
the ordered and unordered model versions share the exact same underlying
generative model, and differ only in the prior probability for the group
affinities, we should not expect any strong general tendency on how many
groups are inferred by either variant: If the network has well-defined
groups, they will be uncovered by either model. Otherwise, if the groups
are not well defined, as is typical for empirical networks that admit
different partitions with similar posterior
probability~\cite{peixoto_revealing_2021}, the most appropriate model
will contribute with a smaller penalty for a subset of them, making them
more likely. Whether the selected partitions have more or fewer groups
will depend on details of the network structure. We show this in
Fig.~\ref{fig:dBdS}, where it can be seen that the difference in
description length between the DC-OSBM and DC-SBM is a relatively
poor predictor of which of them uncovers more groups. The larger
prevalence of networks for which DC-OSBM simultaneously provides a
shorter description length and a larger number of groups when compared
to the DC-SBM is better understood as a characteristic of the network
corpus considered, rather than a necessary outcome of the comparison
between these models.

\section{Conclusion}
\label{sec:conclusion}

We have demonstrated how a Bayesian version of the directed
degree-corrected stochastic block model (DC-SBM)---which is originally
invariant to group label permutations---can be suitably modified
allowing the relative ordering of the group labels to be used to achieve
improved compression whenever the underlying network is embedded in a
one-dimensional latent hierarchy, where most edges tend to follow a
preferred direction. The resulting ordered SBM can be used to infer
latent hierarchies together with arbitrary preferences between groups.

We have investigated how degree correction allows the decoupling from
out-/in-degree imbalance and latent hierarchies, thus removing a source
of conflation that exists in most methods that attempt to rank nodes in
a network.

Furthermore, via model selection we showed how it can be determined if
the ordering is in fact statistically supported, or if a better
description can be obtained with an unordered model. This allows us to
evaluate if the ordering obtained is just the necessary outcome of
constraints we impose during inference, or if they indeed provide a more
plausible description of the data.

It is easy to imagine possible extensions of the ideas presented here
that can reveal more detailed relationships between ranking and
community structure. For example, in our model, lateral edges
(i.e. those that do not involve a difference in rank) can only occur
between nodes of the same group. A potential modification would be to
allow lateral edges between nodes of different groups. Going further, we
could even completely decouple group membership from rank, and infer the
relationship between these properties from the data rather than assume
it a priori --- at the expense of a more complicated model and inference
procedure. We leave such possibilities for future work.

\bibliography{bib,data}

\appendix

\section{The directed, degree-corrected SBM}\label{app:priors}

As shown in the main text, and derived in
Ref.~\cite{peixoto_nonparametric_2017}, the microcanonical directed
degree-corrected SBM has a likelihood given by
\begin{equation}
  P(\A|\bm k, \bm e, \bm b) =
  \frac{\prod_{rs}e_{rs}!\prod_ik_i^{\text{out}}!k_i^{\text{in}}!}
  {\prod_{ij}A_{ij}!\prod_re_r^{\text{out}}!e_r^{\text{in}}!},
\end{equation}
where $\bm k = \{(k_i^{\text{out}},k_i^{\text{in}})\}$ is the imposed
out-/in-degree sequence, with
\begin{equation}
  k_i^{\text{out}} = \sum_j A_{ji},\quad k_i^{\text{in}} = \sum_j A_{ij},
\end{equation}
and $\bm e = \{e_{rs}\}$ being the edge counts between groups, with
marginals given by
\begin{equation}
  e_r^{\text{out}} = \sum_s e_{sr},\quad e_r^{\text{in}} = \sum_s e_{rs}.
\end{equation}
The prior for out-/in-degree sequence is conditioned on the
out-/in-degree distributions
$\bm\eta^{\text{out}}=\{\eta^r_{k^{\text{out}}}\}$ and $\bm\eta^{\text{in}}=\{\eta^r_{k^{\text{in}}}\}$, where
$\eta^r_{k^{\text{out}}}$ ($\eta^r_{k^{\text{in}}}$) is the number of nodes in group
$r$ with out-degree (in-degree) equal to $k^{\text{out}}$ ($k^{\text{in}}$), and
is given by
\begin{equation}
  P(\bm k | \bm \eta^{\text{out}}, \bm \eta^{\text{in}}) = \prod_r \left[\prod_{k^{\text{out}}}\frac{\eta^r_{k^{\text{out}}}!}{n_r!}\right]\left[\prod_{k^{\text{in}}}\frac{\eta^r_{k^{\text{in}}}!}{n_r!}\right],
\end{equation}
with $n_r$ being the number of nodes in group $r$. The out-/in-degree
distributions themselved sampled from group-wise uniform distributions,
\begin{equation}
  P(\bm \eta^{\text{out}}, \bm \eta^{\text{in}} | \bm e, \bb) = \prod_r q(e^{\text{out}}, n_r)^{-1} q(e^{\text{in}}, n_r)^{-1},
\end{equation}
where $q(m,n)=q(m,n-1) +q(m-n,n)$, with boundary conditions $q(m,1)=1$
for $m>0$ and $q(m,n)=0$ for $m\leq 0$ or $n\leq 0$, is the number of
restricted partitions of the integer $m$ into at most $n$ parts.

The non-degree-corrected version of the model can be obtained by
replacing the above prior for $\bm k$ with
\begin{equation}
  P(\bm k | \bm e, \bb) = \prod_r \frac{e^{\text{out}}_r!}{n_r^{e^{\text{out}}}\prod_i(k_i^{\text{out}}!)^{\delta_{b_i,r}}}\times
  \frac{e^{\text{in}}_r!}{n_r^{e^{\text{in}}}\prod_i(k_i^{\text{in}}!)^{\delta_{b_i,r}}}.
\end{equation}
For the partition we have the prior
\begin{align}
  P(\bb) &= P(\bb | \bm n) P(\bm n|B)P(B)\\
         &= \frac{\prod_rn_r!}{N!}{N-1\choose B-1}^{-1}\frac{1}{N}.
\end{align}
Finally, for the edge counts we have a uniform distribution
\begin{equation}
  P(\bm e|E,B) = \multiset{B^2}{E}^{-1},
\end{equation}
or a nested sequence of SBMs for the nested version of the model, as
described in Ref.~\cite{peixoto_nonparametric_2017}.

\section{MCMC inference}\label{app:mcmc}
The inference procedure we use in this work is Markov chain Monte Carlo
(MCMC), implemented as follows. Starting from a partition $\bb$, a new
partition $\bb'$ is proposed with probability $P(\bb'|\bb)$ and accepted
according to the Metropolis-Hastings
criterion~\cite{metropolis_equation_1953,hastings_monte_1970},
i.e. with a probability given by
\begin{equation}
  \min\left(\frac{P(\bb'|\A)P(\bb|\bb')}{P(\bb|\A)P(\bb'|\bb)},1\right),
\end{equation}
otherwise it is rejected. If the move proposals are ergodic and
aperiodic, repeating the above procedure will eventually sample
partitions from the target distribution $P(\bb|\A)$, which needs to be
computed only up to a normalization constant. The move proposals we use
are the merge-split moves described in
Ref.~\cite{peixoto_merge-split_2020} which have very good mixing
properties, and allow each sweep of the algorithm (i.e. a number of
moves that allow each node to change its membership at least once) to be
computed in linear time $O(N+E)$, independent on the number of groups
being considered at any given time.

Although the above method can be used indistinguishably for the ordered
and unordered SBMs, it is beneficial to modify it in a subtle way for
the ordered variant. Since the unordered SBM is invariant to label
permutations, the implementation of the above algorithm can be done
without taking into consideration which labels are used when a new group
is created. On the other hand, with the ordered model, the relative
ordering of the newly created group becomes important. Instead of using
the numeric value of the label itself, it is fact more efficient to
associate with each label $r$ an auxiliary real numeric value $u_r\in
[0, 1]$ which establishes its ordering,
i.e. $r<s$ if and only if $u_r < u_s$. Thus, whenever a new group $r$ is
created, its relative placement is given a new value $u_r$ sampled
uniformly at random in the interval $[0,1]$. The ergodicity of this
auxiliary variable is preserved by allowing the move of the nodes of a
group $r$ to a newly created group $s$, with a new value of $u_s$. In
this way, we can sample re-orderings of the group labels without
actually having to change them.

The above approach will sample partitions from the posterior
distribution. To obtain the partition that maximizes it, we need simply
to add an inverse temperature parameter $\beta$, i.e. $P(\bb|\A) \to
P(\bb|\A)^{\beta}$, and compute the limit $\beta\to\infty$, which means
we only accept a move proposal if it strictly increases the posterior
probability.

A C++ implementation of the above algorithm is available as part of the
\texttt{graph-tool} library~\cite{peixoto_graph-tool_2014}.

\FloatBarrier
\section{Network data}\label{app:data}

In table~\ref{tab:empirical} we list the network data used in this work,
which are freely available from the Netzschleuder
repository~\cite{peixoto_netzschleuder_2020}.

\LTcapwidth=\textwidth
\begin{longtable*}{llS[table-format=5.0]S[table-format=7.0]S[table-format=8.1]S[table-format=8.1]S[table-format=8.1]S[table-format=8.1]l}
  Index & Network & {$N$} & {$E$} & {$\Sigma_{\text{SBM}}$} & {$\Sigma_{\text{DC-SBM}}$}  & {$\Sigma_{\text{OSBM}}$} & {$\Sigma_{\text{DC-OSBM}}$} & Best model\\\hline\\[-.8em]\hline\\[-.4em]
  0 & \href{https://networks.skewed.de/net/genetic_multiplex#Oryctolagus}{\texttt{genetic\_multiplex} (1)}~\cite{domenico_muxviz:_2014} & 79 & 78 & 144.3 & 58.7 & 145.9 & 58.7 & DC-SBM \\
1 & \href{https://networks.skewed.de/net/dom#Shimoji_2014a}{\texttt{dom} (1)}~\cite{strauss_domarchive:_2022} & 20 & 97 & 238.3 & 215 & 235 & 216 & DC-SBM \\
2 & \href{https://networks.skewed.de/net/genetic_multiplex#HepatitusCVirus}{\texttt{genetic\_multiplex} (2)}~\cite{domenico_muxviz:_2014} & 103 & 136 & 288.5 & 229.8 & 290.1 & 231.4 & DC-SBM \\
3 & \href{https://networks.skewed.de/net/genetic_multiplex#DanioRerio}{\texttt{genetic\_multiplex} (3)}~\cite{domenico_muxviz:_2014} & 31 & 54 & 266.3 & 233.8 & 263.3 & 230.3 & DC-OSBM \\
4 & \href{https://networks.skewed.de/net/fresh_webs#Catlins}{\texttt{fresh\_webs} (1)}~\cite{thompson_impacts_2003} & 48 & 110 & 479.8 & 434.2 & 472.9 & 435.7 & DC-SBM \\
5 & \href{https://networks.skewed.de/net/genetic_multiplex#Bos_Multiplex_Genetic}{\texttt{genetic\_multiplex} (4)}~\cite{domenico_muxviz:_2014} & 64 & 74 & 493.6 & 448.3 & 493.9 & 449.7 & DC-SBM \\
6 & \href{https://networks.skewed.de/net/add_health#comm77}{\texttt{add\_health} (1)}~\cite{moody_peer_2001} & 25 & 145 & 488.5 & 469.1 & 483.8 & 468.3 & DC-OSBM \\
7 & \href{https://networks.skewed.de/net/add_health#comm3}{\texttt{add\_health} (2)}~\cite{moody_peer_2001} & 32 & 127 & 515.8 & 487.9 & 514.5 & 486 & DC-OSBM \\
8 & \href{https://networks.skewed.de/net/dom#Shimoji_2014b}{\texttt{dom} (2)}~\cite{strauss_domarchive:_2022} & 32 & 277 & 579.3 & 520.2 & 518.7 & 499.5 & DC-OSBM \\
9 & \href{https://networks.skewed.de/net/fresh_webs#AkatoreB}{\texttt{fresh\_webs} (2)}~\cite{thompson_impacts_2003} & 54 & 117 & 545.1 & 503.7 & 538.3 & 505.2 & DC-SBM \\
10 & \href{https://networks.skewed.de/net/dutch_school#klas12b-net-2}{\texttt{dutch\_school} (1)}~\cite{snijders_introduction_2010} & 26 & 352 & 531.7 & 539.2 & 530.2 & 537.7 & OSBM \\
11 & \href{https://networks.skewed.de/net/genetic_multiplex#HumanHerpes4}{\texttt{genetic\_multiplex} (5)}~\cite{domenico_muxviz:_2014} & 189 & 226 & 748.6 & 564.5 & 779.8 & 573.6 & DC-SBM \\
12 & \href{https://networks.skewed.de/net/fresh_webs#Coweeta1}{\texttt{fresh\_webs} (3)}~\cite{thompson_impacts_2003} & 58 & 126 & 620 & 571.7 & 614.9 & 571.7 & DC-SBM \\
13 & \href{https://networks.skewed.de/net/dutch_school#klas12b-net-4m}{\texttt{dutch\_school} (2)}~\cite{snijders_introduction_2010} & 26 & 578 & 627.4 & 643 & 626.8 & 644.3 & OSBM \\
14 & \href{https://networks.skewed.de/net/dutch_school#klas12b-net-4}{\texttt{dutch\_school} (3)}~\cite{snijders_introduction_2010} & 26 & 629 & 631.3 & 649.3 & 631.1 & 650.6 & OSBM \\
15 & \href{https://networks.skewed.de/net/high_tech_company}{\texttt{high\_tech\_company}}~\cite{krackhardt_cognitive_1987} & 21 & 312 & 680 & 646.8 & 680.5 & 646.8 & DC-SBM \\
16 & \href{https://networks.skewed.de/net/fresh_webs#Coweeta17}{\texttt{fresh\_webs} (4)}~\cite{thompson_impacts_2003} & 71 & 148 & 716.2 & 667.9 & 714 & 667.9 & DC-SBM \\
17 & \href{https://networks.skewed.de/net/fresh_webs#Narrowdale}{\texttt{fresh\_webs} (5)}~\cite{thompson_impacts_2003} & 71 & 155 & 728.2 & 685.6 & 716.3 & 685.6 & DC-SBM \\
18 & \href{https://networks.skewed.de/net/rhesus_monkey}{\texttt{rhesus\_monkey}}~\cite{sade_sociometrics_1972} & 16 & 647 & 720.5 & 715.3 & 710.4 & 704.1 & DC-OSBM \\
19 & \href{https://networks.skewed.de/net/fresh_webs#Venlaw}{\texttt{fresh\_webs} (6)}~\cite{thompson_impacts_2003} & 66 & 187 & 833.8 & 786.5 & 821.5 & 783.6 & DC-OSBM \\
20 & \href{https://networks.skewed.de/net/dom#Hobson_2015b}{\texttt{dom} (3)}~\cite{strauss_domarchive:_2022} & 18 & 810 & 800 & 791.4 & 864.7 & 846.8 & DC-SBM \\
21 & \href{https://networks.skewed.de/net/dutch_school#klas12b-net-3m}{\texttt{dutch\_school} (4)}~\cite{snijders_introduction_2010} & 26 & 1042 & 812.7 & 846.7 & 1033.1 & 933.3 & SBM \\
22 & \href{https://networks.skewed.de/net/dutch_school#klas12b-net-3}{\texttt{dutch\_school} (5)}~\cite{snijders_introduction_2010} & 26 & 1093 & 817 & 853 & 1046 & 945.9 & SBM \\
23 & \href{https://networks.skewed.de/net/fresh_webs#Troy}{\texttt{fresh\_webs} (7)}~\cite{thompson_impacts_2003} & 77 & 181 & 901.8 & 834.8 & 887.9 & 836.3 & DC-SBM \\
24 & \href{https://networks.skewed.de/net/genetic_multiplex#Gallus}{\texttt{genetic\_multiplex} (6)}~\cite{domenico_muxviz:_2014} & 205 & 272 & 1009.8 & 850.3 & 1000.5 & 837 & DC-OSBM \\
25 & \href{https://networks.skewed.de/net/genetic_multiplex#Candida}{\texttt{genetic\_multiplex} (7)}~\cite{domenico_muxviz:_2014} & 303 & 332 & 1247.9 & 906.9 & 1237.6 & 897.9 & DC-OSBM \\
26 & \href{https://networks.skewed.de/net/fresh_webs#AkatoreA}{\texttt{fresh\_webs} (8)}~\cite{thompson_impacts_2003} & 84 & 227 & 1025.7 & 962.9 & 1020 & 964.4 & DC-SBM \\
27 & \href{https://networks.skewed.de/net/fresh_webs#Berwick}{\texttt{fresh\_webs} (9)}~\cite{thompson_impacts_2003} & 77 & 240 & 1033 & 975.7 & 1023.8 & 977.2 & DC-SBM \\
28 & \href{https://networks.skewed.de/net/fresh_webs#NorthCol}{\texttt{fresh\_webs} (10)}~\cite{thompson_impacts_2003} & 78 & 241 & 1036.7 & 1001.2 & 1018.4 & 995.9 & DC-OSBM \\
29 & \href{https://networks.skewed.de/net/add_health#comm76}{\texttt{add\_health} (3)}~\cite{moody_peer_2001} & 43 & 250 & 1050.9 & 1020.3 & 1044.2 & 1010.9 & DC-OSBM \\
30 & \href{https://networks.skewed.de/net/dom#Hobson_2015a}{\texttt{dom} (4)}~\cite{strauss_domarchive:_2022} & 21 & 838 & 1030 & 1022.7 & 1048.6 & 1020.5 & DC-OSBM \\
31 & \href{https://networks.skewed.de/net/hens}{\texttt{hens}}~\cite{guhl_social_1968} & 32 & 496 & 1042.9 & 1022.8 & 1040.2 & 1022.8 & DC-SBM \\
32 & \href{https://networks.skewed.de/net/dom#Shimoji_2014c}{\texttt{dom} (5)}~\cite{strauss_domarchive:_2022} & 48 & 1305 & 1139.6 & 1075.4 & 1095.6 & 1046.9 & DC-OSBM \\
33 & \href{https://networks.skewed.de/net/fresh_webs#Powder}{\texttt{fresh\_webs} (11)}~\cite{thompson_impacts_2003} & 78 & 268 & 1116.1 & 1048.2 & 1103.2 & 1048.4 & DC-SBM \\
34 & \href{https://networks.skewed.de/net/cattle}{\texttt{cattle}}~\cite{schein_social_1955} & 28 & 498 & 1109.8 & 1058.4 & 1059.8 & 1058.7 & DC-SBM \\
35 & \href{https://networks.skewed.de/net/swingers}{\texttt{swingers}}~\cite{niekamp_sexual_2013} & 96 & 232 & 1093.4 & 1063.3 & 1095 & 1063.3 & DC-SBM \\
36 & \href{https://networks.skewed.de/net/dom#Franz_2015b}{\texttt{dom} (6)}~\cite{strauss_domarchive:_2022} & 28 & 1667 & 1244.5 & 1116.3 & 1192 & 1161.2 & DC-SBM \\
37 & \href{https://networks.skewed.de/net/fresh_webs#SuttonAu}{\texttt{fresh\_webs} (12)}~\cite{thompson_impacts_2003} & 80 & 335 & 1237.2 & 1185.2 & 1233.9 & 1181.7 & DC-OSBM \\
38 & \href{https://networks.skewed.de/net/fresh_webs#SuttonSp}{\texttt{fresh\_webs} (13)}~\cite{thompson_impacts_2003} & 74 & 391 & 1230.5 & 1206.9 & 1220.5 & 1208.5 & DC-SBM \\
39 & \href{https://networks.skewed.de/net/fresh_webs#SuttonSu}{\texttt{fresh\_webs} (14)}~\cite{thompson_impacts_2003} & 87 & 843 & 1261.8 & 1229.4 & 1247.2 & 1226.6 & DC-OSBM \\
40 & \href{https://networks.skewed.de/net/moreno_sheep}{\texttt{moreno\_sheep}}~\cite{hass_social_1991} & 28 & 658 & 1287.4 & 1247.8 & 1275.9 & 1250.9 & DC-SBM \\
41 & \href{https://networks.skewed.de/net/fresh_webs#German}{\texttt{fresh\_webs} (15)}~\cite{thompson_impacts_2003} & 84 & 353 & 1344.4 & 1272.7 & 1327.2 & 1263.6 & DC-OSBM \\
42 & \href{https://networks.skewed.de/net/fresh_webs#LilKyeburn}{\texttt{fresh\_webs} (16)}~\cite{thompson_impacts_2003} & 78 & 375 & 1341.7 & 1276.3 & 1331.8 & 1270.3 & DC-OSBM \\
43 & \href{https://networks.skewed.de/net/dom#Foerster_2016a}{\texttt{dom} (7)}~\cite{strauss_domarchive:_2022} & 22 & 2741 & 1322.7 & 1274.3 & 1647.6 & 1539.7 & DC-SBM \\
44 & \href{https://networks.skewed.de/net/7th_graders}{\texttt{7th\_graders}}~\cite{mathews_secondary_1976} & 29 & 740 & 1356.1 & 1358.8 & 1349.2 & 1348.2 & DC-OSBM \\
45 & \href{https://networks.skewed.de/net/fresh_webs#DempstersAu}{\texttt{fresh\_webs} (17)}~\cite{thompson_impacts_2003} & 83 & 415 & 1441.3 & 1360.9 & 1436.1 & 1354.7 & DC-OSBM \\
46 & \href{https://networks.skewed.de/net/add_health#comm1}{\texttt{add\_health} (4)}~\cite{moody_peer_2001} & 69 & 305 & 1421.5 & 1377.9 & 1398.4 & 1361.6 & DC-OSBM \\
47 & \href{https://networks.skewed.de/net/fresh_webs#Blackrock}{\texttt{fresh\_webs} (18)}~\cite{thompson_impacts_2003} & 86 & 375 & 1468.8 & 1412.5 & 1455.4 & 1410.7 & DC-OSBM \\
48 & \href{https://networks.skewed.de/net/bison}{\texttt{bison}}~\cite{lott_dominance_2010} & 26 & 897 & 1506.8 & 1460 & 1474.8 & 1443.6 & DC-OSBM \\
49 & \href{https://networks.skewed.de/net/software_dependencies#junit}{\texttt{software\_dependencies} (1)}~\cite{vsubelj_software_2012,vsubelj_community_2011,vsubelj_software_2012,lovro_clustering_2012,vsubelj_node_2014} & 105 & 451 & 1664.6 & 1519.5 & 1626.1 & 1485.8 & DC-OSBM \\
50 & \href{https://networks.skewed.de/net/fresh_webs#Martins}{\texttt{fresh\_webs} (19)}~\cite{thompson_impacts_2003} & 105 & 343 & 1600.4 & 1521.8 & 1581.6 & 1510.2 & DC-OSBM \\
51 & \href{https://networks.skewed.de/net/dom#Foerster_2016b}{\texttt{dom} (8)}~\cite{strauss_domarchive:_2022} & 44 & 1015 & 1787 & 1622.9 & 1749.5 & 1646.5 & DC-SBM \\
52 & \href{https://networks.skewed.de/net/sp_baboons#observational}{\texttt{sp\_baboons} (1)}~\cite{gelardi_measuring_2020} & 23 & 3197 & 1738.3 & 1725.1 & 1657.5 & 1653.1 & DC-OSBM \\
53 & \href{https://networks.skewed.de/net/fresh_webs#Broad}{\texttt{fresh\_webs} (20)}~\cite{thompson_impacts_2003} & 94 & 565 & 1828.8 & 1742.6 & 1822.7 & 1737.8 & DC-OSBM \\
54 & \href{https://networks.skewed.de/net/fresh_webs#DempstersSp}{\texttt{fresh\_webs} (21)}~\cite{thompson_impacts_2003} & 93 & 538 & 1834.6 & 1745.7 & 1822.5 & 1741.1 & DC-OSBM \\
55 & \href{https://networks.skewed.de/net/add_health#comm63}{\texttt{add\_health} (5)}~\cite{moody_peer_2001} & 96 & 352 & 1917.6 & 1793.5 & 1898.3 & 1771.4 & DC-OSBM \\
56 & \href{https://networks.skewed.de/net/highschool}{\texttt{highschool}}~\cite{coleman1964introduction} & 70 & 506 & 1931.5 & 1886.4 & 1892.2 & 1854.3 & DC-OSBM \\
57 & \href{https://networks.skewed.de/net/add_health#comm70}{\texttt{add\_health} (6)}~\cite{moody_peer_2001} & 76 & 440 & 1977.9 & 1874.9 & 1948.3 & 1858.4 & DC-OSBM \\
58 & \href{https://networks.skewed.de/net/add_health#comm71}{\texttt{add\_health} (7)}~\cite{moody_peer_2001} & 74 & 466 & 1969.5 & 1887.6 & 1952.8 & 1875.2 & DC-OSBM \\
59 & \href{https://networks.skewed.de/net/dom#Franz_2015c}{\texttt{dom} (9)}~\cite{strauss_domarchive:_2022} & 36 & 2387 & 2065.1 & 1905 & 2153.5 & 2060.4 & DC-SBM \\
60 & \href{https://networks.skewed.de/net/fresh_webs#Kyeburn}{\texttt{fresh\_webs} (22)}~\cite{thompson_impacts_2003} & 98 & 629 & 2072.4 & 1984.8 & 2053.6 & 1983 & DC-OSBM \\
61 & \href{https://networks.skewed.de/net/fresh_webs#Healy}{\texttt{fresh\_webs} (23)}~\cite{thompson_impacts_2003} & 96 & 634 & 2144 & 2040.1 & 2128.6 & 2033.5 & DC-OSBM \\
62 & \href{https://networks.skewed.de/net/add_health#comm2}{\texttt{add\_health} (8)}~\cite{moody_peer_2001} & 103 & 445 & 2285.5 & 2177.9 & 2252.8 & 2160.6 & DC-OSBM \\
63 & \href{https://networks.skewed.de/net/add_health#comm6}{\texttt{add\_health} (9)}~\cite{moody_peer_2001} & 108 & 457 & 2396.2 & 2268 & 2353.2 & 2255.9 & DC-OSBM \\
64 & \href{https://networks.skewed.de/net/kidnappings}{\texttt{kidnappings}}~\cite{gerdes_assessing_2014} & 285 & 357 & 2499.5 & 2382.2 & 2501.1 & 2376.9 & DC-OSBM \\
65 & \href{https://networks.skewed.de/net/fresh_webs#Canton}{\texttt{fresh\_webs} (24)}~\cite{thompson_impacts_2003} & 109 & 875 & 2492.9 & 2406.8 & 2479.3 & 2387.5 & DC-OSBM \\
66 & \href{https://networks.skewed.de/net/fresh_webs#Stony}{\texttt{fresh\_webs} (25)}~\cite{thompson_impacts_2003} & 112 & 832 & 2561.5 & 2467.6 & 2546 & 2456.4 & DC-OSBM \\
67 & \href{https://networks.skewed.de/net/dom#Franz_2015e}{\texttt{dom} (10)}~\cite{strauss_domarchive:_2022} & 52 & 3281 & 2984.5 & 2752.2 & 3037.2 & 2912.1 & DC-SBM \\
68 & \href{https://networks.skewed.de/net/dom#Franz_2015d}{\texttt{dom} (11)}~\cite{strauss_domarchive:_2022} & 53 & 4464 & 3069 & 2843.3 & 3199 & 3041.1 & DC-SBM \\
69 & \href{https://networks.skewed.de/net/fresh_webs#DempstersSu}{\texttt{fresh\_webs} (26)}~\cite{thompson_impacts_2003} & 107 & 966 & 2997.8 & 2851.8 & 2977.3 & 2844.8 & DC-OSBM \\
70 & \href{https://networks.skewed.de/net/genetic_multiplex#Xenopus}{\texttt{genetic\_multiplex} (8)}~\cite{domenico_muxviz:_2014} & 263 & 427 & 3223.4 & 2925 & 3161.9 & 2883.9 & DC-OSBM \\
71 & \href{https://networks.skewed.de/net/college_freshmen}{\texttt{college\_freshmen}}~\cite{bunt__1999} & 32 & 3062 & 2936.8 & 2885.1 & 2960.5 & 2920.4 & DC-SBM \\
72 & \href{https://networks.skewed.de/net/physician_trust}{\texttt{physician\_trust}}~\cite{coleman_diffusion_1957} & 117 & 542 & 3191.6 & 3056.5 & 3175.3 & 3045.8 & DC-OSBM \\
73 & \href{https://networks.skewed.de/net/freshmen#t0}{\texttt{freshmen} (1)}~\cite{bunt__1999} & 34 & 6908 & 3054.8 & 3118.6 & 3054.8 & 3118.6 & SBM \\
74 & \href{https://networks.skewed.de/net/freshmen#t3}{\texttt{freshmen} (2)}~\cite{bunt__1999} & 34 & 5781 & 3199.7 & 3219.2 & 3201.2 & 3219.2 & SBM \\
75 & \href{https://networks.skewed.de/net/freshmen#t6}{\texttt{freshmen} (3)}~\cite{bunt__1999} & 34 & 6484 & 3213.3 & 3331.6 & 3282.9 & 3350.1 & SBM \\
76 & \href{https://networks.skewed.de/net/ecoli_transcription#v1.1}{\texttt{ecoli\_transcription} (1)}~\cite{shen-orr_network_2002} & 328 & 497 & 3511.1 & 3238.1 & 3465.9 & 3216.5 & DC-OSBM \\
77 & \href{https://networks.skewed.de/net/ecoli_transcription#v1.0}{\texttt{ecoli\_transcription} (2)}~\cite{shen-orr_network_2002} & 329 & 496 & 3519 & 3241.5 & 3474.3 & 3226.5 & DC-OSBM \\
78 & \href{https://networks.skewed.de/net/freshmen#t2}{\texttt{freshmen} (4)}~\cite{bunt__1999} & 34 & 6009 & 3247.2 & 3288.4 & 3246.9 & 3288.4 & OSBM \\
79 & \href{https://networks.skewed.de/net/freshmen#t5}{\texttt{freshmen} (5)}~\cite{bunt__1999} & 34 & 6492 & 3255.2 & 3312.9 & 3273 & 3312.9 & SBM \\
80 & \href{https://networks.skewed.de/net/software_dependencies#jmail}{\texttt{software\_dependencies} (2)}~\cite{vsubelj_software_2012,vsubelj_community_2011,vsubelj_software_2012,lovro_clustering_2012,vsubelj_node_2014} & 192 & 875 & 3772.4 & 3415.6 & 3660.1 & 3367.9 & DC-OSBM \\
81 & \href{https://networks.skewed.de/net/software_dependencies#flamingo}{\texttt{software\_dependencies} (3)}~\cite{vsubelj_software_2012,vsubelj_community_2011,vsubelj_software_2012,lovro_clustering_2012,vsubelj_node_2014} & 228 & 813 & 4068.4 & 3665.5 & 3969.6 & 3589.3 & DC-OSBM \\
82 & \href{https://networks.skewed.de/net/dom#Franz_2015a}{\texttt{dom} (12)}~\cite{strauss_domarchive:_2022} & 61 & 4118 & 3991.4 & 3738.1 & 4100.5 & 3897.4 & DC-SBM \\
83 & \href{https://networks.skewed.de/net/macaques}{\texttt{macaques}}~\cite{fedigan1991monkeys} & 62 & 2435 & 4844.4 & 4777.4 & 4756.1 & 4687.3 & DC-OSBM \\
84 & \href{https://networks.skewed.de/net/add_health#comm5}{\texttt{add\_health} (10)}~\cite{moody_peer_2001} & 157 & 945 & 4975.9 & 4862.8 & 4932.9 & 4834 & DC-OSBM \\
85 & \href{https://networks.skewed.de/net/genetic_multiplex#HumanHIV1}{\texttt{genetic\_multiplex} (9)}~\cite{domenico_muxviz:_2014} & 1005 & 1355 & 6240 & 5417.7 & 6306 & 5441.5 & DC-SBM \\
86 & \href{https://networks.skewed.de/net/add_health#comm8}{\texttt{add\_health} (11)}~\cite{moody_peer_2001} & 204 & 1012 & 5775.8 & 5558.7 & 5718.5 & 5506.8 & DC-OSBM \\
87 & \href{https://networks.skewed.de/net/software_dependencies#sjbullet}{\texttt{software\_dependencies} (4)}~\cite{vsubelj_software_2012,vsubelj_community_2011,vsubelj_software_2012,lovro_clustering_2012,vsubelj_node_2014} & 249 & 1726 & 6298 & 5682.3 & 6138.1 & 5611.5 & DC-OSBM \\
88 & \href{https://networks.skewed.de/net/law_firm}{\texttt{law\_firm}}~\cite{raub_emmanuel_2005} & 71 & 2571 & 6075.9 & 5881.6 & 5959.7 & 5845.5 & DC-OSBM \\
89 & \href{https://networks.skewed.de/net/foodweb_little_rock}{\texttt{foodweb\_little\_rock}}~\cite{martinez_artifacts_1991} & 183 & 2494 & 6195.8 & 6105.3 & 6167.2 & 6090.6 & DC-OSBM \\
90 & \href{https://networks.skewed.de/net/foodweb_baywet}{\texttt{foodweb\_baywet}}~\cite{ulanowicz1999network} & 128 & 2106 & 6437.4 & 6351.8 & 6423.5 & 6358.6 & DC-SBM \\
91 & \href{https://networks.skewed.de/net/add_health#comm37}{\texttt{add\_health} (12)}~\cite{moody_peer_2001} & 358 & 869 & 6820.5 & 6595.3 & 6740.4 & 6516.2 & DC-OSBM \\
92 & \href{https://networks.skewed.de/net/software_dependencies#jung}{\texttt{software\_dependencies} (5)}~\cite{vsubelj_software_2012,vsubelj_community_2011,vsubelj_software_2012,lovro_clustering_2012,vsubelj_node_2014} & 398 & 1716 & 7510.2 & 6856.7 & 7303.6 & 6729.6 & DC-OSBM \\
93 & \href{https://networks.skewed.de/net/software_dependencies#guava}{\texttt{software\_dependencies} (6)}~\cite{vsubelj_software_2012,vsubelj_community_2011,vsubelj_software_2012,lovro_clustering_2012,vsubelj_node_2014} & 457 & 2668 & 7557.4 & 6852 & 7421.9 & 6763.9 & DC-OSBM \\
94 & \href{https://networks.skewed.de/net/software_dependencies#sjung}{\texttt{software\_dependencies} (7)}~\cite{vsubelj_software_2012,vsubelj_community_2011,vsubelj_software_2012,lovro_clustering_2012,vsubelj_node_2014} & 399 & 1721 & 7583.9 & 6895.5 & 7323 & 6764.8 & DC-OSBM \\
95 & \href{https://networks.skewed.de/net/add_health#comm55}{\texttt{add\_health} (13)}~\cite{moody_peer_2001} & 331 & 1006 & 7193.2 & 6939.7 & 7081 & 6848.2 & DC-OSBM \\
96 & \href{https://networks.skewed.de/net/add_health#comm9}{\texttt{add\_health} (14)}~\cite{moody_peer_2001} & 248 & 1264 & 7312 & 7051.6 & 7207.3 & 6963.9 & DC-OSBM \\
97 & \href{https://networks.skewed.de/net/yeast_transcription}{\texttt{yeast\_transcription}}~\cite{milo_network_2002} & 664 & 1078 & 7838.5 & 7362.2 & 7773 & 7321.4 & DC-OSBM \\
98 & \href{https://networks.skewed.de/net/software_dependencies#colt}{\texttt{software\_dependencies} (8)}~\cite{vsubelj_software_2012,vsubelj_community_2011,vsubelj_software_2012,lovro_clustering_2012,vsubelj_node_2014} & 504 & 3677 & 8371.3 & 7666.6 & 8256.3 & 7457.2 & DC-OSBM \\
99 & \href{https://networks.skewed.de/net/software_dependencies#scolt}{\texttt{software\_dependencies} (9)}~\cite{vsubelj_software_2012,vsubelj_community_2011,vsubelj_software_2012,lovro_clustering_2012,vsubelj_node_2014} & 504 & 3677 & 8335.3 & 7637.2 & 8206.2 & 7459.3 & DC-OSBM \\
100 & \href{https://networks.skewed.de/net/add_health#comm67}{\texttt{add\_health} (15)}~\cite{moody_peer_2001} & 439 & 1065 & 8293.2 & 7851.4 & 8191 & 7752.7 & DC-OSBM \\
101 & \href{https://networks.skewed.de/net/add_health#comm4}{\texttt{add\_health} (16)}~\cite{moody_peer_2001} & 281 & 1396 & 8585.2 & 8225.8 & 8515.6 & 8192.7 & DC-OSBM \\
102 & \href{https://networks.skewed.de/net/add_health#comm18}{\texttt{add\_health} (17)}~\cite{moody_peer_2001} & 284 & 1511 & 9513.8 & 9222.3 & 9390 & 9122.4 & DC-OSBM \\
103 & \href{https://networks.skewed.de/net/add_health#comm72}{\texttt{add\_health} (18)}~\cite{moody_peer_2001} & 352 & 1784 & 10564.9 & 10112.1 & 10442.7 & 10003.3 & DC-OSBM \\
104 & \href{https://networks.skewed.de/net/faculty_hiring#history}{\texttt{faculty\_hiring} (1)}~\cite{clauset_systematic_2015} & 144 & 4112 & 10923 & 10516.1 & 11037.8 & 10537 & DC-SBM \\
105 & \href{https://networks.skewed.de/net/add_health#comm56}{\texttt{add\_health} (19)}~\cite{moody_peer_2001} & 444 & 1652 & 11739.9 & 11210.5 & 11550.2 & 11109.7 & DC-OSBM \\
106 & \href{https://networks.skewed.de/net/add_health#comm78}{\texttt{add\_health} (20)}~\cite{moody_peer_2001} & 430 & 1718 & 11714.1 & 11339.7 & 11529.4 & 11188.9 & DC-OSBM \\
107 & \href{https://networks.skewed.de/net/add_health#comm21}{\texttt{add\_health} (21)}~\cite{moody_peer_2001} & 377 & 2021 & 12101.6 & 11795.6 & 11932.5 & 11634.4 & DC-OSBM \\
108 & \href{https://networks.skewed.de/net/cintestinalis}{\texttt{cintestinalis}}~\cite{ryan_cns_2016} & 205 & 2903 & 12122.3 & 11754 & 12181.9 & 11787.7 & DC-SBM \\
109 & \href{https://networks.skewed.de/net/celegansneural}{\texttt{celegansneural}}~\cite{structure_1986,watts_collective_1998} & 297 & 2359 & 12060.8 & 11814.8 & 12063.5 & 11798 & DC-OSBM \\
110 & \href{https://networks.skewed.de/net/software_dependencies#org}{\texttt{software\_dependencies} (10)}~\cite{vsubelj_software_2012,vsubelj_community_2011,vsubelj_software_2012,lovro_clustering_2012,vsubelj_node_2014} & 486 & 4990 & 13207.6 & 12495.1 & 13066.7 & 12395.9 & DC-OSBM \\
111 & \href{https://networks.skewed.de/net/add_health#comm11}{\texttt{add\_health} (22)}~\cite{moody_peer_2001} & 411 & 1975 & 12991.7 & 12539.5 & 12865.8 & 12455.3 & DC-OSBM \\
112 & \href{https://networks.skewed.de/net/add_health#comm53}{\texttt{add\_health} (23)}~\cite{moody_peer_2001} & 579 & 1814 & 13636.2 & 13033.2 & 13514.8 & 12913.9 & DC-OSBM \\
113 & \href{https://networks.skewed.de/net/dom#Strauss_2019d}{\texttt{dom} (13)}~\cite{strauss_domarchive:_2022} & 151 & 9096 & 13763.9 & 13469.3 & 13432.5 & 13252 & DC-OSBM \\
114 & \href{https://networks.skewed.de/net/add_health#comm7}{\texttt{add\_health} (24)}~\cite{moody_peer_2001} & 437 & 2155 & 13927.4 & 13541.3 & 13746.9 & 13381.6 & DC-OSBM \\
115 & \href{https://networks.skewed.de/net/faculty_hiring#business}{\texttt{faculty\_hiring} (2)}~\cite{clauset_systematic_2015} & 112 & 7856 & 14622 & 14136.3 & 15061.4 & 14310.5 & DC-SBM \\
116 & \href{https://networks.skewed.de/net/add_health#comm31}{\texttt{add\_health} (25)}~\cite{moody_peer_2001} & 728 & 2012 & 15413.5 & 14702.3 & 15266.3 & 14509.3 & DC-OSBM \\
117 & \href{https://networks.skewed.de/net/faculty_hiring#computer_science}{\texttt{faculty\_hiring} (3)}~\cite{clauset_systematic_2015} & 205 & 4388 & 15247.2 & 14762.8 & 15337.2 & 14725.4 & DC-OSBM \\
118 & \href{https://networks.skewed.de/net/software_dependencies#jung-c}{\texttt{software\_dependencies} (11)}~\cite{vsubelj_software_2012,vsubelj_community_2011,vsubelj_software_2012,lovro_clustering_2012,vsubelj_node_2014} & 879 & 5339 & 16733.1 & 15224.4 & 16252.4 & 14837 & DC-OSBM \\
119 & \href{https://networks.skewed.de/net/add_health#comm51}{\texttt{add\_health} (26)}~\cite{moody_peer_2001} & 676 & 1949 & 15628.4 & 14969 & 15500.7 & 14858.7 & DC-OSBM \\
120 & \href{https://networks.skewed.de/net/add_health#comm80}{\texttt{add\_health} (27)}~\cite{moody_peer_2001} & 594 & 2188 & 15837.5 & 15153.6 & 15633.1 & 15006.6 & DC-OSBM \\
121 & \href{https://networks.skewed.de/net/add_health#comm74}{\texttt{add\_health} (28)}~\cite{moody_peer_2001} & 654 & 2064 & 16246.7 & 15499.3 & 16021.5 & 15355.6 & DC-OSBM \\
122 & \href{https://networks.skewed.de/net/add_health#comm26}{\texttt{add\_health} (29)}~\cite{moody_peer_2001} & 551 & 2624 & 16443.4 & 15748.5 & 16232.5 & 15591.9 & DC-OSBM \\
123 & \href{https://networks.skewed.de/net/add_health#comm65}{\texttt{add\_health} (30)}~\cite{moody_peer_2001} & 557 & 2327 & 16439.6 & 16023.3 & 16263.4 & 15878.6 & DC-OSBM \\
124 & \href{https://networks.skewed.de/net/add_health#comm38}{\texttt{add\_health} (31)}~\cite{moody_peer_2001} & 521 & 2340 & 16735.7 & 16180.5 & 16598.1 & 16098.9 & DC-OSBM \\
125 & \href{https://networks.skewed.de/net/celegans_2019#male_chemical_synapse}{\texttt{celegans\_2019} (1)}~\cite{cook_whole-animal_2019} & 328 & 3531 & 16986.5 & 16328.4 & 16775.3 & 16243.2 & DC-OSBM \\
126 & \href{https://networks.skewed.de/net/add_health#comm19}{\texttt{add\_health} (32)}~\cite{moody_peer_2001} & 492 & 2675 & 17225.9 & 16739.2 & 16969.2 & 16529.5 & DC-OSBM \\
127 & \href{https://networks.skewed.de/net/celegans_2019#hermaphrodite_chemical_synapse}{\texttt{celegans\_2019} (2)}~\cite{cook_whole-animal_2019} & 313 & 3534 & 17045.7 & 16697.9 & 17015.3 & 16653.1 & DC-OSBM \\
128 & \href{https://networks.skewed.de/net/add_health#comm22}{\texttt{add\_health} (33)}~\cite{moody_peer_2001} & 612 & 3132 & 19786.3 & 19115.7 & 19496.2 & 18872.9 & DC-OSBM \\
129 & \href{https://networks.skewed.de/net/add_health#comm29}{\texttt{add\_health} (34)}~\cite{moody_peer_2001} & 569 & 3203 & 20018.9 & 19434.7 & 19725.8 & 19216.1 & DC-OSBM \\
130 & \href{https://networks.skewed.de/net/add_health#comm13}{\texttt{add\_health} (35)}~\cite{moody_peer_2001} & 652 & 2935 & 20602.4 & 19814 & 20397.6 & 19645.3 & DC-OSBM \\
131 & \href{https://networks.skewed.de/net/add_health#comm14}{\texttt{add\_health} (36)}~\cite{moody_peer_2001} & 562 & 3344 & 21289.1 & 20572.8 & 21085.9 & 20381.4 & DC-OSBM \\
132 & \href{https://networks.skewed.de/net/add_health#comm12}{\texttt{add\_health} (37)}~\cite{moody_peer_2001} & 581 & 3585 & 22262.5 & 21641.5 & 21965.6 & 21376.1 & DC-OSBM \\
133 & \href{https://networks.skewed.de/net/add_health#comm10}{\texttt{add\_health} (38)}~\cite{moody_peer_2001} & 678 & 3441 & 22795 & 21888.8 & 22570.4 & 21704.8 & DC-OSBM \\
134 & \href{https://networks.skewed.de/net/residence_hall}{\texttt{residence\_hall}}~\cite{freeman_exploring_1998} & 217 & 9028 & 24200.8 & 24021.6 & 22474 & 22541.4 & OSBM \\
135 & \href{https://networks.skewed.de/net/genetic_multiplex#Plasmodium}{\texttt{genetic\_multiplex} (10)}~\cite{domenico_muxviz:_2014} & 1158 & 2497 & 23403.9 & 22494.9 & 23456.1 & 22481.8 & DC-OSBM \\
136 & \href{https://networks.skewed.de/net/add_health#comm25}{\texttt{add\_health} (39)}~\cite{moody_peer_2001} & 790 & 3178 & 23765.8 & 22783.5 & 23490.1 & 22520.5 & DC-OSBM \\
137 & \href{https://networks.skewed.de/net/celegans_2019#hermaphrodite_chemical_corrected}{\texttt{celegans\_2019} (3)}~\cite{cook_whole-animal_2019} & 446 & 4879 & 23328.7 & 23101.7 & 23573.4 & 23092.2 & DC-OSBM \\
138 & \href{https://networks.skewed.de/net/celegans_2019#hermaphrodite_chemical}{\texttt{celegans\_2019} (4)}~\cite{cook_whole-animal_2019} & 446 & 4879 & 23307.7 & 23097.3 & 23525.6 & 23132.2 & DC-SBM \\
139 & \href{https://networks.skewed.de/net/add_health#comm30}{\texttt{add\_health} (40)}~\cite{moody_peer_2001} & 718 & 3442 & 24778.2 & 23946.9 & 24428.6 & 23696.5 & DC-OSBM \\
140 & \href{https://networks.skewed.de/net/add_health#comm66}{\texttt{add\_health} (41)}~\cite{moody_peer_2001} & 644 & 3591 & 24628.7 & 23944.7 & 24335.8 & 23728.5 & DC-OSBM \\
141 & \href{https://networks.skewed.de/net/add_health#comm23}{\texttt{add\_health} (42)}~\cite{moody_peer_2001} & 667 & 3783 & 24955.6 & 24237.4 & 24592.4 & 23963.4 & DC-OSBM \\
142 & \href{https://networks.skewed.de/net/add_health#comm64}{\texttt{add\_health} (43)}~\cite{moody_peer_2001} & 694 & 3544 & 25398.2 & 24603.5 & 25182.8 & 24393.2 & DC-OSBM \\
143 & \href{https://networks.skewed.de/net/add_health#comm45}{\texttt{add\_health} (44)}~\cite{moody_peer_2001} & 921 & 3223 & 25872.9 & 24948.3 & 25801.7 & 24828.1 & DC-OSBM \\
144 & \href{https://networks.skewed.de/net/add_health#comm24}{\texttt{add\_health} (45)}~\cite{moody_peer_2001} & 849 & 3735 & 26328.1 & 25313.5 & 26017.6 & 25097.5 & DC-OSBM \\
145 & \href{https://networks.skewed.de/net/celegans_2019#male_chemical_corrected}{\texttt{celegans\_2019} (5)}~\cite{cook_whole-animal_2019} & 559 & 5306 & 26051.5 & 25510.4 & 26375.1 & 25615.6 & DC-SBM \\
146 & \href{https://networks.skewed.de/net/celegans_2019#male_chemical}{\texttt{celegans\_2019} (6)}~\cite{cook_whole-animal_2019} & 559 & 5306 & 26024.5 & 25542.2 & 26300.1 & 25585.1 & DC-SBM \\
147 & \href{https://networks.skewed.de/net/add_health#comm62}{\texttt{add\_health} (46)}~\cite{moody_peer_2001} & 1040 & 3321 & 27305.3 & 26319 & 27039.7 & 26041.8 & DC-OSBM \\
148 & \href{https://networks.skewed.de/net/add_health#comm27}{\texttt{add\_health} (47)}~\cite{moody_peer_2001} & 1152 & 3291 & 27243.4 & 26662.8 & 26821.4 & 26250.7 & DC-OSBM \\
149 & \href{https://networks.skewed.de/net/add_health#comm16}{\texttt{add\_health} (48)}~\cite{moody_peer_2001} & 778 & 4125 & 28107.7 & 27176.6 & 28027.9 & 27086.7 & DC-OSBM \\
150 & \href{https://networks.skewed.de/net/add_health#comm35}{\texttt{add\_health} (49)}~\cite{moody_peer_2001} & 851 & 3735 & 29457.2 & 28352.4 & 29252.8 & 28213 & DC-OSBM \\
151 & \href{https://networks.skewed.de/net/add_health#comm54}{\texttt{add\_health} (50)}~\cite{moody_peer_2001} & 1035 & 3710 & 29594 & 28476.3 & 29364.3 & 28234.7 & DC-OSBM \\
152 & \href{https://networks.skewed.de/net/software_dependencies#weka}{\texttt{software\_dependencies} (12)}~\cite{vsubelj_software_2012,vsubelj_community_2011,vsubelj_software_2012,lovro_clustering_2012,vsubelj_node_2014} & 1225 & 9553 & 31018.1 & 28900.7 & 30693.2 & 28550.7 & DC-OSBM \\
153 & \href{https://networks.skewed.de/net/genetic_multiplex#Rattus}{\texttt{genetic\_multiplex} (11)}~\cite{domenico_muxviz:_2014} & 2350 & 4014 & 32290.2 & 29697 & 31990 & 29539.3 & DC-OSBM \\
154 & \href{https://networks.skewed.de/net/add_health#comm59}{\texttt{add\_health} (51)}~\cite{moody_peer_2001} & 971 & 4156 & 31234.1 & 30135.6 & 30875.3 & 29819.2 & DC-OSBM \\
155 & \href{https://networks.skewed.de/net/add_health#comm32}{\texttt{add\_health} (52)}~\cite{moody_peer_2001} & 853 & 4191 & 31290.8 & 30274.7 & 30931.5 & 30008.4 & DC-OSBM \\
156 & \href{https://networks.skewed.de/net/add_health#comm69}{\texttt{add\_health} (53)}~\cite{moody_peer_2001} & 891 & 4561 & 32994.9 & 32070 & 32636.5 & 31821.4 & DC-OSBM \\
157 & \href{https://networks.skewed.de/net/add_health#comm57}{\texttt{add\_health} (54)}~\cite{moody_peer_2001} & 1180 & 4282 & 34045.8 & 32902.2 & 33643.7 & 32575.8 & DC-OSBM \\
158 & \href{https://networks.skewed.de/net/messal_shale}{\texttt{messal\_shale}}~\cite{dunne_highly_2014} & 700 & 6444 & 33400.8 & 32694.3 & 33384.4 & 32688.5 & DC-OSBM \\
159 & \href{https://networks.skewed.de/net/add_health#comm60}{\texttt{add\_health} (55)}~\cite{moody_peer_2001} & 1131 & 4684 & 34566.7 & 33331.4 & 34084.4 & 32930.3 & DC-OSBM \\
160 & \href{https://networks.skewed.de/net/add_health#comm20}{\texttt{add\_health} (56)}~\cite{moody_peer_2001} & 910 & 5229 & 36392.8 & 35564.9 & 35965.5 & 35192.4 & DC-OSBM \\
161 & \href{https://networks.skewed.de/net/add_health#comm83}{\texttt{add\_health} (57)}~\cite{moody_peer_2001} & 1260 & 4520 & 37160.4 & 35722.2 & 36628.3 & 35353.2 & DC-OSBM \\
162 & \href{https://networks.skewed.de/net/add_health#comm39}{\texttt{add\_health} (58)}~\cite{moody_peer_2001} & 987 & 4881 & 37011.6 & 35884.5 & 36608.3 & 35594.1 & DC-OSBM \\
163 & \href{https://networks.skewed.de/net/add_health#comm82}{\texttt{add\_health} (59)}~\cite{moody_peer_2001} & 921 & 5094 & 37626.3 & 36623.8 & 37210.3 & 36295.7 & DC-OSBM \\
164 & \href{https://networks.skewed.de/net/email_company}{\texttt{email\_company}}~\cite{michalski_matching_2011} & 167 & 82927 & 38086.2 & 37879.3 & 37033.3 & 36905.2 & DC-OSBM \\
165 & \href{https://networks.skewed.de/net/add_health#comm75}{\texttt{add\_health} (60)}~\cite{moody_peer_2001} & 994 & 5459 & 38842.3 & 37626.6 & 38432.4 & 37354.9 & DC-OSBM \\
166 & \href{https://networks.skewed.de/net/add_health#comm47}{\texttt{add\_health} (61)}~\cite{moody_peer_2001} & 985 & 5410 & 39106.5 & 38033.3 & 38775.4 & 37732.1 & DC-OSBM \\
167 & \href{https://networks.skewed.de/net/add_health#comm68}{\texttt{add\_health} (62)}~\cite{moody_peer_2001} & 1385 & 4845 & 39624.3 & 38411.7 & 39114.7 & 37995.8 & DC-OSBM \\
168 & \href{https://networks.skewed.de/net/add_health#comm81}{\texttt{add\_health} (63)}~\cite{moody_peer_2001} & 1290 & 4689 & 39940.1 & 38528.1 & 39519.1 & 38085.9 & DC-OSBM \\
169 & \href{https://networks.skewed.de/net/add_health#comm15}{\texttt{add\_health} (64)}~\cite{moody_peer_2001} & 1062 & 5370 & 40111.6 & 38809.6 & 39739.2 & 38448.3 & DC-OSBM \\
170 & \href{https://networks.skewed.de/net/add_health#comm84}{\texttt{add\_health} (65)}~\cite{moody_peer_2001} & 1545 & 4775 & 40906.3 & 39124.1 & 40561.8 & 38813.4 & DC-OSBM \\
171 & \href{https://networks.skewed.de/net/add_health#comm79}{\texttt{add\_health} (66)}~\cite{moody_peer_2001} & 1190 & 5371 & 41657 & 40540.6 & 41174.7 & 40037.1 & DC-OSBM \\
172 & \href{https://networks.skewed.de/net/interactome_figeys}{\texttt{interactome\_figeys}}~\cite{ewing_large-scale_2007} & 2217 & 6438 & 41729.3 & 40612.4 & 41798.6 & 40546 & DC-OSBM \\
173 & \href{https://networks.skewed.de/net/interactome_stelzl}{\texttt{interactome\_stelzl}}~\cite{stelzl_human_2005} & 1615 & 6105 & 45150.1 & 42551.5 & 43105.2 & 41239.6 & DC-OSBM \\
174 & \href{https://networks.skewed.de/net/add_health#comm28}{\texttt{add\_health} (67)}~\cite{moody_peer_2001} & 1136 & 5720 & 42972.9 & 41644.3 & 42624.9 & 41304.5 & DC-OSBM \\
175 & \href{https://networks.skewed.de/net/software_dependencies#javax}{\texttt{software\_dependencies} (13)}~\cite{vsubelj_software_2012,vsubelj_community_2011,vsubelj_software_2012,lovro_clustering_2012,vsubelj_node_2014} & 1570 & 17273 & 45177.9 & 41693.2 & 44567.2 & 41882.4 & DC-SBM \\
176 & \href{https://networks.skewed.de/net/add_health#comm44}{\texttt{add\_health} (68)}~\cite{moody_peer_2001} & 1127 & 6189 & 44766.2 & 43575.2 & 44442.2 & 43367.2 & DC-OSBM \\
177 & \href{https://networks.skewed.de/net/add_health#comm61}{\texttt{add\_health} (69)}~\cite{moody_peer_2001} & 1710 & 5380 & 46082.7 & 44428.1 & 45662.4 & 43953 & DC-OSBM \\
178 & \href{https://networks.skewed.de/net/add_health#comm48}{\texttt{add\_health} (70)}~\cite{moody_peer_2001} & 1171 & 6217 & 45615.8 & 44291.1 & 45125.7 & 43998.6 & DC-OSBM \\
179 & \href{https://networks.skewed.de/net/add_health#comm42}{\texttt{add\_health} (71)}~\cite{moody_peer_2001} & 1405 & 5621 & 48146 & 46573.2 & 48026.1 & 46463.5 & DC-OSBM \\
180 & \href{https://networks.skewed.de/net/add_health#comm17}{\texttt{add\_health} (72)}~\cite{moody_peer_2001} & 1218 & 6488 & 50698.2 & 49069 & 50106 & 48597.3 & DC-OSBM \\
181 & \href{https://networks.skewed.de/net/add_health#comm43}{\texttt{add\_health} (73)}~\cite{moody_peer_2001} & 1638 & 6339 & 54897.9 & 53339.3 & 54579.8 & 52877 & DC-OSBM \\
182 & \href{https://networks.skewed.de/net/add_health#comm58}{\texttt{add\_health} (74)}~\cite{moody_peer_2001} & 1703 & 7015 & 55615.8 & 53760.8 & 55021.1 & 53277.2 & DC-OSBM \\
183 & \href{https://networks.skewed.de/net/add_health#comm33}{\texttt{add\_health} (75)}~\cite{moody_peer_2001} & 1974 & 5849 & 56030.6 & 54067.7 & 55571.5 & 53480.8 & DC-OSBM \\
184 & \href{https://networks.skewed.de/net/add_health#comm52}{\texttt{add\_health} (76)}~\cite{moody_peer_2001} & 1719 & 6772 & 55937.1 & 53989.9 & 55285.5 & 53496.9 & DC-OSBM \\
185 & \href{https://networks.skewed.de/net/add_health#comm34}{\texttt{add\_health} (77)}~\cite{moody_peer_2001} & 1605 & 6984 & 58002.5 & 56114.4 & 57490.6 & 55638.9 & DC-OSBM \\
186 & \href{https://networks.skewed.de/net/add_health#comm46}{\texttt{add\_health} (78)}~\cite{moody_peer_2001} & 1519 & 7149 & 59212.8 & 57543.5 & 58710.3 & 56906.2 & DC-OSBM \\
187 & \href{https://networks.skewed.de/net/word_adjacency#japanese}{\texttt{word\_adjacency} (1)}~\cite{milo_superfamilies_2004} & 2698 & 8297 & 59830 & 57085.2 & 60715 & 58249.2 & DC-SBM \\
188 & \href{https://networks.skewed.de/net/uni_email}{\texttt{uni\_email}}~\cite{guimera_self-similar_2003} & 1133 & 10903 & 65340 & 62916.4 & 61968.1 & 60652.5 & DC-OSBM \\
189 & \href{https://networks.skewed.de/net/add_health#comm73}{\texttt{add\_health} (79)}~\cite{moody_peer_2001} & 1630 & 8556 & 65754.8 & 63908.4 & 64956.8 & 63252.6 & DC-OSBM \\
190 & \href{https://networks.skewed.de/net/software_dependencies#slucene}{\texttt{software\_dependencies} (14)}~\cite{vsubelj_software_2012,vsubelj_community_2011,vsubelj_software_2012,lovro_clustering_2012,vsubelj_node_2014} & 2811 & 17373 & 70635 & 65903.9 & 70183.5 & 65465.7 & DC-OSBM \\
191 & \href{https://networks.skewed.de/net/add_health#comm49}{\texttt{add\_health} (80)}~\cite{moody_peer_2001} & 1877 & 8869 & 68314.2 & 66318.9 & 67548.6 & 65659.8 & DC-OSBM \\
192 & \href{https://networks.skewed.de/net/genetic_multiplex#Celegans}{\texttt{genetic\_multiplex} (12)}~\cite{domenico_muxviz:_2014} & 3692 & 8058 & 69341.1 & 66202.8 & 69421.2 & 66008.1 & DC-OSBM \\
193 & \href{https://networks.skewed.de/net/software_dependencies#java}{\texttt{software\_dependencies} (15)}~\cite{vsubelj_software_2012,vsubelj_community_2011,vsubelj_software_2012,lovro_clustering_2012,vsubelj_node_2014} & 2378 & 34858 & 78653.7 & 73278.1 & 79145.7 & 73971.2 & DC-SBM \\
194 & \href{https://networks.skewed.de/net/add_health#comm36}{\texttt{add\_health} (81)}~\cite{moody_peer_2001} & 2152 & 9878 & 79277.9 & 76641.5 & 78374.7 & 75867.8 & DC-OSBM \\
195 & \href{https://networks.skewed.de/net/add_health#comm40}{\texttt{add\_health} (82)}~\cite{moody_peer_2001} & 1996 & 10485 & 84806.2 & 82781.3 & 84027.1 & 81842.1 & DC-OSBM \\
196 & \href{https://networks.skewed.de/net/fao_trade}{\texttt{fao\_trade}}~\cite{domenico_structural_2015} & 214 & 318346 & 81952.6 & 82042.9 & 117912.1 & 104149.6 & SBM \\
197 & \href{https://networks.skewed.de/net/add_health#comm41}{\texttt{add\_health} (83)}~\cite{moody_peer_2001} & 2064 & 10503 & 85277.1 & 82662 & 84598.7 & 81966.2 & DC-OSBM \\
198 & \href{https://networks.skewed.de/net/polblogs}{\texttt{polblogs}}~\cite{adamic_political_2005} & 1222 & 19089 & 89057.8 & 84867.2 & 90735.5 & 85165.9 & DC-SBM \\
199 & \href{https://networks.skewed.de/net/add_health#comm50}{\texttt{add\_health} (84)}~\cite{moody_peer_2001} & 2539 & 12969 & 109236.4 & 106320.9 & 107934.2 & 105453.8 & DC-OSBM \\
200 & \href{https://networks.skewed.de/net/genetic_multiplex#Arabidopsis}{\texttt{genetic\_multiplex} (13)}~\cite{domenico_muxviz:_2014} & 6692 & 18397 & 152555.6 & 144702.5 & 152892.7 & 143924 & DC-OSBM \\
201 & \href{https://networks.skewed.de/net/genetic_multiplex#Mus}{\texttt{genetic\_multiplex} (14)}~\cite{domenico_muxviz:_2014} & 7402 & 19553 & 174481.5 & 164850.6 & 174312.8 & 164965.1 & DC-SBM \\
202 & \href{https://networks.skewed.de/net/word_adjacency#french}{\texttt{word\_adjacency} (2)}~\cite{milo_superfamilies_2004} & 8308 & 24286 & 199836.5 & 190855.8 & 204061.3 & 195700.2 & DC-SBM \\
203 & \href{https://networks.skewed.de/net/gnutella#08}{\texttt{gnutella} (1)}~\cite{ripeanu_mapping_2002} & 6299 & 20776 & 210066.1 & 205119.1 & 210909.1 & 205949.4 & DC-SBM \\
204 & \href{https://networks.skewed.de/net/jung}{\texttt{jung}}~\cite{Subelj_software_2012} & 6120 & 138706 & 253033.7 & 239074.1 & 256683 & 243697.4 & DC-SBM \\
205 & \href{https://networks.skewed.de/net/software_dependencies#jung-j}{\texttt{software\_dependencies} (16)}~\cite{vsubelj_software_2012,vsubelj_community_2011,vsubelj_software_2012,lovro_clustering_2012,vsubelj_node_2014} & 6120 & 138706 & 253610.5 & 239652.3 & 255954.9 & 243249.6 & DC-SBM \\
206 & \href{https://networks.skewed.de/net/software_dependencies#jdk}{\texttt{software\_dependencies} (17)}~\cite{vsubelj_software_2012,vsubelj_community_2011,vsubelj_software_2012,lovro_clustering_2012,vsubelj_node_2014} & 6434 & 150985 & 267968 & 254399 & 271869.4 & 258955 & DC-SBM \\
207 & \href{https://networks.skewed.de/net/jdk}{\texttt{jdk}}~\cite{kunegis_konect_2013} & 6434 & 150985 & 267842 & 254631.1 & 272583.1 & 257675.3 & DC-SBM \\
208 & \href{https://networks.skewed.de/net/gnutella#09}{\texttt{gnutella} (2)}~\cite{ripeanu_mapping_2002} & 8104 & 26008 & 274397.7 & 267235.4 & 275039.2 & 268022.8 & DC-SBM \\
209 & \href{https://networks.skewed.de/net/word_adjacency#darwin}{\texttt{word\_adjacency} (3)}~\cite{milo_superfamilies_2004} & 7377 & 46279 & 307720 & 297896 & 316867.3 & 307916.4 & DC-SBM \\
210 & \href{https://networks.skewed.de/net/word_adjacency#spanish}{\texttt{word\_adjacency} (4)}~\cite{milo_superfamilies_2004} & 11558 & 45114 & 317093.8 & 305261.6 & 323260.4 & 311244.2 & DC-SBM \\
211 & \href{https://networks.skewed.de/net/advogato}{\texttt{advogato}}~\cite{massa_bowling_2009} & 5042 & 49631 & 342611.1 & 325964.7 & 340306.7 & 323051.8 & DC-OSBM \\
212 & \href{https://networks.skewed.de/net/gnutella#06}{\texttt{gnutella} (3)}~\cite{ripeanu_mapping_2002} & 8717 & 31525 & 347389.6 & 338865.6 & 349573.1 & 340673.8 & DC-SBM \\
213 & \href{https://networks.skewed.de/net/genetic_multiplex#Sacchpomb}{\texttt{genetic\_multiplex} (15)}~\cite{domenico_muxviz:_2014} & 4078 & 63667 & 350660.4 & 341319 & 360342.5 & 344945.4 & DC-SBM \\
214 & \href{https://networks.skewed.de/net/genetic_multiplex#Drosophila}{\texttt{genetic\_multiplex} (16)}~\cite{domenico_muxviz:_2014} & 8114 & 43304 & 379841 & 366866.3 & 379162.4 & 365044.6 & DC-OSBM \\
215 & \href{https://networks.skewed.de/net/dblp_cite}{\texttt{dblp\_cite}}~\cite{ley_dblp_2002} & 12494 & 49702 & 424041.2 & 399517 & 427323.1 & 400700.4 & DC-SBM \\
216 & \href{https://networks.skewed.de/net/inploid}{\texttt{inploid}}~\cite{gursoy_influence_2018} & 14360 & 57101 & 426144.2 & 408307.9 & 428266.2 & 408654.5 & DC-SBM \\
217 & \href{https://networks.skewed.de/net/anybeat}{\texttt{anybeat}}~\cite{fire_link_2012} & 12645 & 67053 & 442601.1 & 418413.9 & 440836.3 & 419607.8 & DC-SBM \\
218 & \href{https://networks.skewed.de/net/gnutella#04}{\texttt{gnutella} (4)}~\cite{ripeanu_mapping_2002} & 10876 & 39994 & 461535.8 & 450107.1 & 463416.7 & 451601.6 & DC-SBM \\
219 & \href{https://networks.skewed.de/net/chess}{\texttt{chess}}~\cite{chess} & 7115 & 64926 & 488399.4 & 476033 & 480542.8 & 469920.4 & DC-OSBM \\
220 & \href{https://networks.skewed.de/net/elec}{\texttt{elec}}~\cite{Leskovec_2010} & 7066 & 103645 & 572402.8 & 545967.9 & 586592.3 & 554748.5 & DC-SBM \\
221 & \href{https://networks.skewed.de/net/caida_as#20071112}{\texttt{caida\_as} (1)}~\cite{caida_as} & 26389 & 105722 & 736834.6 & 683345 & 710941.9 & 659739.5 & DC-OSBM \\
222 & \href{https://networks.skewed.de/net/python_dependency}{\texttt{python\_dependency}}~\cite{python_dependency} & 58302 & 108118 & 723383.2 & 660414.2 & 724114 & 663173.3 & DC-SBM \\
223 & \href{https://networks.skewed.de/net/google}{\texttt{google}}~\cite{palla_directed_2007} & 15763 & 171206 & 697162.3 & 673705.9 & 710447.1 & 685376.1 & DC-SBM \\
224 & \href{https://networks.skewed.de/net/gnutella#25}{\texttt{gnutella} (5)}~\cite{ripeanu_mapping_2002} & 22663 & 54693 & 704797.8 & 687626.8 & 707145.2 & 686795.9 & DC-OSBM \\
225 & \href{https://networks.skewed.de/net/cora}{\texttt{cora}}~\cite{mccallum__2000} & 23166 & 91500 & 854972.6 & 791698.1 & 849458.3 & 787856.1 & DC-OSBM \\
226 & \href{https://networks.skewed.de/net/nematode_mammal}{\texttt{nematode\_mammal}}~\cite{dallas_gauging_2018} & 26197 & 140432 & 816953.4 & 811996.1 & 809148.3 & 805070.2 & DC-OSBM \\
227 & \href{https://networks.skewed.de/net/gnutella#24}{\texttt{gnutella} (6)}~\cite{ripeanu_mapping_2002} & 26498 & 65359 & 857564.7 & 837415.1 & 861399.6 & 836664 & DC-OSBM \\
228 & \href{https://networks.skewed.de/net/fediverse}{\texttt{fediverse}}~\cite{rochko_2018} & 4860 & 484164 & 1062391.4 & 1052004.3 & 1180717.8 & 1077672.5 & DC-SBM \\
229 & \href{https://networks.skewed.de/net/gnutella#30}{\texttt{gnutella} (7)}~\cite{ripeanu_mapping_2002} & 36646 & 88303 & 1169393.7 & 1138636.4 & 1176941.9 & 1144731.6 & DC-SBM \\
230 & \href{https://networks.skewed.de/net/genetic_multiplex#Homo}{\texttt{genetic\_multiplex} (17)}~\cite{domenico_muxviz:_2014} & 18136 & 170831 & 1218093.4 & 1168743.5 & 1222213.7 & 1171322.5 & DC-SBM \\
231 & \href{https://networks.skewed.de/net/linux}{\texttt{linux}}~\cite{kunegis_konect_2013} & 30817 & 213942 & 1332386.6 & 1288094.1 & 1349663 & 1294157.3 & DC-SBM \\
232 & \href{https://networks.skewed.de/net/genetic_multiplex#Sacchcere}{\texttt{genetic\_multiplex} (18)}~\cite{domenico_muxviz:_2014} & 6567 & 282752 & 1584214 & 1534637.8 & 1590960 & 1537713.9 & DC-SBM \\
233 & \href{https://networks.skewed.de/net/pgp_strong}{\texttt{pgp\_strong}}~\cite{richters_trust_2011} & 39796 & 301498 & 1885884.9 & 1780603.7 & 1816232.7 & 1731146.7 & DC-OSBM \\
234 & \href{https://networks.skewed.de/net/scotus_majority#2008}{\texttt{scotus\_majority} (1)}~\cite{fowler_authority_2008,fowler_network_2007} & 25389 & 216718 & 1933664.5 & 1825534.5 & 1957209.5 & 1840510.1 & DC-SBM \\
235 & \href{https://networks.skewed.de/net/scotus_majority#2007}{\texttt{scotus\_majority} (2)}~\cite{fowler_authority_2008,fowler_network_2007} & 34428 & 202053 & 2016896.9 & 1903877.7 & 2036012.7 & 1914565.6 & DC-SBM \\
236 & \href{https://networks.skewed.de/net/email_enron}{\texttt{email\_enron}}~\cite{klimt_enron_2004} & 33696 & 361622 & 2272843.4 & 2138109.2 & 2059220.9 & 1997285.2 & DC-OSBM \\
237 & \href{https://networks.skewed.de/net/gnutella#31}{\texttt{gnutella} (8)}~\cite{ripeanu_mapping_2002} & 62561 & 147878 & 2083997.2 & 2029112.1 & 2094327.1 & 2036521.2 & DC-SBM \\
238 & \href{https://networks.skewed.de/net/arxiv_citation#HepTh}{\texttt{arxiv\_citation} (1)}~\cite{gehrke_overview_2003} & 27400 & 352542 & 2391491 & 2219924.9 & 2409700.1 & 2258057.8 & DC-SBM \\
239 & \href{https://networks.skewed.de/net/arxiv_citation#HepPh}{\texttt{arxiv\_citation} (2)}~\cite{gehrke_overview_2003} & 34401 & 421485 & 3005722.6 & 2792846.7 & 3017691.7 & 2812687.5 & DC-SBM \\
240 & \href{https://networks.skewed.de/net/email_eu}{\texttt{email\_eu}}~\cite{leskovec_graph_2007} & 224832 & 395270 & 3598338.7 & 3356699.3 & 3631807 & 3358836 & DC-SBM \\
241 & \href{https://networks.skewed.de/net/word_assoc}{\texttt{word\_assoc}}~\cite{kiss1973associative} & 23132 & 511764 & 3722536.2 & 3529298.9 & 3746228 & 3507729.5 & DC-OSBM \\
242 & \href{https://networks.skewed.de/net/facebook_wall}{\texttt{facebook\_wall}}~\cite{viswanath_evolution_2009} & 43953 & 872044 & 4751164.5 & 4217967.1 & 4296512 & 3922272.1 & DC-OSBM \\
243 & \href{https://networks.skewed.de/net/epinions_trust}{\texttt{epinions\_trust}}~\cite{richardson_trust_2003} & 75877 & 508836 & 4521591 & 4341144.1 & 4491015.6 & 4325874.1 & DC-OSBM \\
244 & \href{https://networks.skewed.de/net/notre_dame_web}{\texttt{notre\_dame\_web}}~\cite{albert_diameter_1999} & 325729 & 1497134 & 9206468.8 & 8520817.4 & 9221048.2 & 8516072.9 & DC-OSBM \\
245 & \href{https://networks.skewed.de/net/stanford_web}{\texttt{stanford\_web}}~\cite{jure_community_2008} & 255265 & 2234572 & 10013497.4 & 9532902.7 & 10221145.9 & 9718934.4 & DC-SBM \\
246 & \href{https://networks.skewed.de/net/google_plus}{\texttt{google\_plus}}~\cite{fire_computationally_2013} & 201949 & 1496936 & 10452997.8 & 10043644.6 & 10411311.3 & 9923644.7 & DC-OSBM \\
247 & \href{https://networks.skewed.de/net/genetic_multiplex#YeastLandscape}{\texttt{genetic\_multiplex} (19)}~\cite{domenico_muxviz:_2014} & 4458 & 8473997 & 13093737.9 & 13111684.6 & 13425751.3 & 13431929.3 & SBM \\
248 & \href{https://networks.skewed.de/net/academia_edu}{\texttt{academia\_edu}}~\cite{fire_computationally_2013} & 200167 & 1398062 & 14600464.4 & 13643254.2 & 14168442.1 & 13156559.8 & DC-OSBM \\
249 & \href{https://networks.skewed.de/net/citeseer}{\texttt{citeseer}}~\cite{bollacker_citeseer_1998} & 365154 & 1736325 & 19208311.4 & 17476771.6 & 18825114.3 & 17422382.8 & DC-OSBM \\
250 & \href{https://networks.skewed.de/net/berkstan_web}{\texttt{berkstan\_web}}~\cite{jure_community_2008} & 654782 & 7499425 & 30088771 & 28559020.8 & 30068489.5 & 28621191.2 & DC-SBM
  \\\hline\\

  \caption{Directed network data used in this work, indexed in
  increasing order of minimum description length (in accordance with
  Fig.~\ref{fig:empirical}), together with the number of nodes $N$ and
  edges $E$, the description length in bits obtained with the four model
  variants, as well as the model with the shortest description
  length. \label{tab:empirical}}
\end{longtable*}

\end{document}